\documentclass[a4paper,11pt]{article}
\pdfoutput=1 
\usepackage{jheppub} 
\usepackage{bbm,bm,graphicx,mathtools,color,hyperref,slashed}
\usepackage{amssymb,amsmath}
\usepackage[T1]{fontenc}

\usepackage[all]{xy}

\newcommand{\diff}{\mathrm{d}}
\newcommand{\p}{\partial}
\newcommand{\ve}{\varepsilon}

\newcommand{\Diff}{{\mathcal{D}}}

\newcommand{\be}{\begin{equation}}      
\newcommand{\ee}{\end{equation}}      
\newcommand{\bea}{\begin{eqnarray}}      
\newcommand{\eea}{\end{eqnarray}}

\newcommand{\tr}{\mathrm{tr}}

\newcommand{\im}{\mathrm{i}}

\newcommand{\calC}{\mathcal{C}}

\newcommand{\calI}{\mathcal{I}}
\newcommand{\calO}{\mathcal{O}}

\newcommand{\calT}{\mathcal{T}}

\newcommand{\rme}{\mathrm{e}}

\newcommand{\bphi}{\bm{\phi}}
\newcommand{\with}{\quad\mathrm{with}\quad}
\newcommand{\at}{\quad\mathrm{at}\quad}
\newcommand{\ol}[1]{{\overline{#1}}}

\newcommand{\tilbphi}{\widetilde{\bm{\phi}}}
\newcommand{\tilphi}{\widetilde{\phi}}

\newcommand{\bra}[1]{\langle {#1} |}

\newcommand{\ket}[1]{| {#1} \rangle}

\title{
Phase structure of the twisted $SU(3)/U(1)^2$ flag sigma model on $\mathbb{R}\times S^1$
}

\author[a]{Masaru Hongo,}
\author[b,a,c]{Tatsuhiro Misumi,}
\author[d]{and Yuya Tanizaki}

\affiliation[a]{iTHEMS Program, RIKEN, Wako 351-0198, Japan}
\affiliation[b]{Department of Mathematical Science, Akita University, Akita 010-8502, Japan}
\affiliation[c]{Research and Education Center for Natural Sciences, Keio University, Kanagawa 223-8521, Japan}
\affiliation[d]{RIKEN BNL Research Center, Brookhaven National Laboratory, Upton, NY 11973 USA}

\emailAdd{masaru.hongo@riken.jp}
\emailAdd{misumi@phys.akita-u.ac.jp}
\emailAdd{yuya.tanizaki@riken.jp}

\abstract{
We investigate the phase structure of the compactified $2$-dimensional 
nonlinear $SU(3)/U(1)^2$ flag sigma model with respect to two $\theta$-terms.
Based on the circle compactification with the ${\mathbb Z}_{3}$-twisted boundary condition, 
which preserves an 't~Hooft anomaly of the original uncompactified theory, 
we perform the semiclassical analysis based on the dilute instanton gas approximation (DIGA).
We clarify classical vacua of the theory and derive fractional instanton solutions connecting these vacua.
The resulting phase structure based on DIGA exhibits the quantum phase transitions 
and triple degeneracy at special points in the $(\theta_1,\theta_2)$-plane,
which is consistent with the phase diagram obtained from the anomaly matching and 
global inconsistency conditions.
This result indicates the adiabatic continuity between the flag sigma models on ${\mathbb R}^{2}$ and ${\mathbb R}\times S^{1}$ with small compactification radius.
We further estimate contributions from instanton--anti-instanton configuration (bion) and show
the existence of the imaginary ambiguity, which is expected to be cancelled by that of the perturbative Borel resummation.
}

\begin{document}
\maketitle

%---------------------------------------------------------------------------------------
\section{Introduction}
\label{sec:Intro}

$2$-dimensional nonlinear sigma models with the target space $SU(N)/U(1)^{N-1}$ (flag sigma models) have been recently getting much attention since it gives the low-energy effective description of $SU(N)$ anti-ferromagnetic spin chains~\cite{Bykov:2011ai, Lajko:2017wif}. 
This generalizes the Haldane conjecture by finding its connection with $SU(N)$ Wess-Zumino-Witten model~\cite{Tanizaki:2018xto, Ohmori:2018qza}, which is also supported by the Lieb-Schultz-Mattis theorem~\cite{Lieb:1961fr, Affleck:1986pq,PhysRevLett.84.1535} or 't~Hooft anomaly matching condition~\cite{tHooft:1979rat, Frishman:1980dq} (see also Refs.~\cite{Wen:2013oza, Wang:2014pma, Wang:2014tia, Tachikawa:2016cha, Gaiotto:2017yup, Tanizaki:2017bam, Komargodski:2017dmc, Komargodski:2017smk, Tanizaki:2017qhf, Tanizaki:2017mtm, Cherman:2017dwt, Sulejmanpasic:2018upi, Yao:2018kel, Anber:2018iof, Cordova:2018acb, Anber:2018jdf, Tanizaki:2018wtg} for recent developments). 
%Nonperturbative properties of the flag sigma models have been investigated in terms of the 't Hooft anomaly matching and the global inconsistency~\cite{Tanizaki:2018xto, Ohmori:2018qza}, where the phase diagram with respect to the multiple $\theta$ angles is conjectured.
Phase diagram of the flag sigma model with respect to multiple $\theta$ angles has been studied by the lattice strong-coupling expansion~\cite{Lajko:2017wif}, and also by the anomaly matching and global inconsistency condition~\cite{Tanizaki:2018xto}. 
Since this theory has asymptotic freedom, however, the model cannot be studied reliably based on the semiclassical analysis in two dimensions~\cite{Witten:1978bc, Affleck:1979gy}. This prevents us from verifying the conjectured phase diagram by the use of the semiclassical analysis with instantons. 

Recently, adiabatic continuity under circle compactification has been proposed for asymptotically-free field theories including adjoint QCD~\cite{Unsal:2007vu, Kovtun:2007py, Unsal:2007jx, Unsal:2008ch, Shifman:2008ja, Shifman:2009tp} and $2$-dimensional nonlinear sigma models~\cite{Dunne:2016nmc, Dunne:2012ae}. For nonlinear sigma models, this roughly states that the phase structure of the theory on $\mathbb{R}\times S^1$ with a symmetry-twisted boundary condition coincides with that of the original $2$-dimensional theory. This property is still at the level of conjecture in general cases, but has been proved for large-$N$ limit of $\mathbb{C}P^{N-1}$ and $O(N)$ sigma models~\cite{Sulejmanpasic:2016llc}.
Adiabatic continuity implies that one can investigate the phase diagram of asymptotically-free field theories based on semiclassical analysis with small compactification circumference of $S^{1}$.
In the ${\mathbb C}P^{N-1}$ and $SU(N)/U(1)^{N-1}$ flag sigma models on ${\mathbb R} \times S^{1}$ with the ${\mathbb Z}_{N}$-twisted boundary condition, the continuity has been also argued in terms of the 't Hooft anomaly matching~\cite{Tanizaki:2017qhf,Tanizaki:2018xto}: the mixed 't Hooft anomaly involving the ${\mathbb Z}_{N}$ shift-center symmetry associated with the ${\mathbb Z}_{N}$-twisted boundary condition survives even in the compactified limit. 

Research on the adiabatic circle compactification is also of great influence on the attempt to give nonperturbative definition of $2$-dimensional field theories based on the resurgence theory~\cite{Dunne:2012ae,Dunne:2012zk,Cherman:2013yfa,Cherman:2014ofa,Misumi:2014jua,Misumi:2014bsa,Dunne:2015ywa,Buividovich:2015oju,Demulder:2016mja,Misumi:2016fno,Fujimori:2016ljw,Fujimori:2017oab,Fujimori:2017osz,Dorigoni:2017smz,Okuyama:2018clk,Fujimori:2018kqp}. Resurgence theory, which was originally developed for the ordinary differential equation, has been applied to a broad area of theoretical physics including, quantum mechanics~\cite{Lipatov:1977cd,ZinnJustin:1981dx,Aoyama:1997qk,Alvarez1,ZinnJustin:2004ib,Dunne:2013ada,Basar:2013eka,Misumi:2015dua,Behtash:2015zha,Kozcaz:2016wvy}, string theory~\cite{Marino:2006hs,Marino:2007te,Marino:2008ya,Marino:2008vx,Aniceto:2011nu,Schiappa:2013opa,Aniceto:2013fka,Santamaria:2013rua,Grassi:2014cla,Couso-Santamaria:2014iia,Couso-Santamaria:2015wga,Aniceto:2015rua,Dorigoni:2015dha,Kuroki:2016ucm,Couso-Santamaria:2016vwq,Arutyunov:2016etw} and supersymmetric field theory~\cite{Aniceto:2014hoa,Honda:2016mvg,Honda:2016vmv,Honda:2017qdb,Fujimori:2018nvz} (see more references in \cite{Marino:2012zq,Aniceto:2018bis}). 
%When trying to give the definition of asymptotically-free theories in the continuum spacetime, we encounter the problem of renormalon singularities at the nonperturbative level. 
Since the application of resurgence theory to quantum theories utilizes the nontrivial relation between perturbative and nonperturbative contributions to physical quantities, the adiabatic continuity between the perturbative (weak-coupling) and nonperturbative (strong-coupling) parameter regions are essential for continuing results to the nonperturbative region.
Turning our eyes to the $SU(N)/U(1)^{N-1}$ flag sigma models with these facts in mind, what we have to do is to figure out whether the adiabatic continuity exists and the nonperturbative properties including the phase structure are maintained in the process of compactification.

In this work, we study the phase diagram in the parameter space of the two theta angles ($\theta_{1},\theta_{2}$) in the $SU(3)/U(1)^{2}$ flag sigma model on $\mathbb{R}\times S^1$ with the ${\mathbb Z}_{3}$-twisted boundary condition given in Ref.~\cite{Tanizaki:2018xto}. To apply the semiclassical analysis or the dilute instanton gas approximation (DIGA) to the theory, we work on the small compactification circumference in most parts of this work. 

We first classify the classical vacua and derive the fractional instanton solutions of $SU(3)/U(1)^2$ sigma model on $\mathbb{R}\times S^1$ with the ${\mathbb Z}_{3}$-twisted boundary condition. Since $SU(3)/U(1)^2$ sigma model can be regarded as a set of three copies of ${\mathbb C}P^{2}$ sigma models with orthogonality constraints, we can use the knowledge of fractional instantons of $\mathbb{C}P^{N-1}$ model~\cite{Eto:2004rz,Eto:2006mz,Eto:2006pg,Bruckmann:2007zh,Brendel:2009mp}. 
We find six independent classical vacua that are connected by eighteen fractional instanton solutions. 
Based on the fractional solutions and classification of the vacua, we obtain the partition function and the six lowest eigenenergies within the DIGA. 
We find the existence of phase transitions in the $(\theta_{1},\theta_{2})$-plane caused by the degeneracy of ground states. 
The resultant phase diagram is in good agreement with the one conjectured by the 't Hooft anomaly matching and the global inconsistency. 
We also show how the 't~Hooft anomaly and the global inconsistency are realized in the approximated Hilbert space in DIGA.
This is consistent with the adiabatic continuity of the phase diagram between strong-coupling (large compactified radius) and weak-coupling (small compactified radius) regions.

We also set about investigation on the resurgent structure of the present theories. We obtain contributions to the partition function from the instanton-antiinstanton configuration called a bion, and show the existence of the imaginary ambiguity by performing the integral with respect to the parameters called quasi-moduli. Although we do not incorporate the one-loop determinant including the quantum fluctuations on the top of bion configurations in this work, the imaginary ambiguity with taking them into account is expected to be cancelled by the perturbative Borel resummation if there is the nontrivial resurgent structure.  

The paper is constructed as follows:
In Sec.~\ref{sec:Model} we discuss the $2$-dimensional $SU(3)/U(1)^2$ sigma model 
and its circle-compactified version with the ${\mathbb Z}_{3}$-twisted boundary condition, with emphasis on the symmetries of the original uncompactified and compactified theories. 
We there argue the conjectured phase diagram and the adiabatic continuity in terms of 't~Hooft anomaly matching.
In Sec.~\ref{sec:PhaseDiagram}, by classifying the classical vacua and deriving the fractional instanton solutions, we obtain the partition function and the eigenenergies within the dilute instanton gas approximation, which leads to the phase diagram in the $(\theta_{1},\theta_{2})$ parameter space.
In Sec.~\ref{sec:Bion}, we calculate contributions from the instanton-antiinstanton configuration and show the existence of the imaginary ambiguity.
Sec.~\ref{sec:Summary} is devoted to the summary and discussion.

%------------------------------------------------------------------------------- --------
\section{$2$d $SU(3)/U(1)^2$ sigma model with twisted boundary condition}
\label{sec:Model}
In this section, we review the $2$-dimensional $SU(3)/U(1)^2$ sigma model 
and its circle compactification with a twisted boundary condition~\cite{Tanizaki:2018xto}. 
We put emphasis on the symmetry structure of the original uncompactified 
and compactified theory, and propose the adiabatic continuity 
between them based on the 't~Hooft anomaly~\cite{Tanizaki:2017qhf, Tanizaki:2017mtm, Dunne:2018hog, Tanizaki:2018xto}.
After explaining the original uncompactified theory in Sec.~\ref{sec:Original},
we introduce the compactified theory, adiabatic continuity conjecture, and 
its symmetry structure in Sec.~\ref{sec:Twisted}.

\subsection{$2$d $SU(3)/U(1)^2$ nonlinear sigma model}
\label{sec:Original}
Let us consider the $2$-dimensional $SU(3)/U(1)^2$ nonlinear sigma model, 
whose action is given by
\begin{equation}
 S [\bphi_\ell,a_\ell] = \sum_{\ell=1}^{3} \int_{M_2} 
  \left[ -{1\over 2 g} |(\diff+\im a_\ell) \bphi_\ell |^2
   +{\im \theta_\ell\over 2\pi} \diff a_\ell 
   +{\lambda\over 2\pi}(\overline{\bphi}_{\ell+1} \cdot \diff \bphi_\ell) 
   \wedge (\bphi_{\ell+1}\cdot \diff \overline{\bphi}_\ell)\right],
  \label{eq:Action}
\end{equation}
where 
$\bm{\phi}_{\ell}=(\phi_{1,\ell},\phi_{2,\ell},\phi_{3,\ell})^t \in \mathbb{C}^3
~(\ell=1,2,3)$ denote three-component complex scalar fields 
interacting with $U(1)$ gauge fields $a_\ell~(\ell=1,2,3)$.
Here $\bphi_\ell$ satisfy
the following constraint:
\begin{equation}
 \overline{\bm{\phi}}_\ell\cdot \bm{\phi}_k=\delta_{\ell k}, 
  \quad
  \ve_{abc} \phi_{a,1} \phi_{b,2} \phi_{c,3} = 1.
\label{eq:constraint_1}
\end{equation}
Introducing the $3\times 3$ matrix made of $\bphi_\ell$,
\begin{equation}
 \Phi=[\bm{\phi}_1,\bm{\phi}_2,\bm{\phi}_3]=
  \begin{pmatrix}
   \phi_{1,1}& \phi_{1,2} & \phi_{1,3}\\
   \phi_{2,1}& \phi_{2,2} & \phi_{2,3}\\
   \phi_{3,1}& \phi_{3,2} & \phi_{3,3}
  \end{pmatrix}, 
\end{equation}
the constraint (\ref{eq:constraint_1}) means $\Phi \in SU(3)$, i.e.
\begin{equation}
 \Phi^\dagger \Phi=\bm{1}_3 , 
  \quad 
  \det \Phi = 1.
  \label{eq:SU(3)}
\end{equation}
Since $\det \Phi$ has charge $1$ under three $U(1)$ gauge symmetries,  there are two independent $U(1)$ gauge fields among $a_\ell~(\ell=1,2,3)$, and 
the target space of this sigma model is given by the flag manifold $SU(3)/U(1)^2$.

The action~\eqref{eq:Action} gives the low-energy effective field theory
of $SU(3)$ anti-ferromagnetic spin chains derived in Ref.~\cite{Lajko:2017wif}.
In addition to the familiar $\theta$ terms, the term proportional to $\lambda$ 
is also topological---but not quantized---in the sence 
that it does not depend on the metric of the base manifold $M_2$.
In this paper, we will mainly focus on the case with vanishing $\lambda$-terms, 
and put $\lambda=0$.
One may na\"ively think that the model reduces to the simple sum of 
three copies of $\mathbb{C}P^2$ nonlinear sigma model when we 
turn off the $\lambda$-term.
However, this is not true because three copies are still coupled 
due to the constraint \eqref{eq:SU(3)}.

We then discuss a consequence coming from the constraints by using the 
equation of motion for $a_\ell$.
As is usual for $\mathbb{C}P^{N-1}$ sigma model, 
only the first term in the action \eqref{eq:Action}---the kinetic term 
for $\bphi_\ell$ and gauge interactions between $\bphi_\ell$ 
and $a_\ell$---contributes to the equation of motion for $a_\ell$, 
which leads to
\begin{equation}
 a_\ell= \im \overline{\bm{\phi}}_\ell\cdot \diff \bm{\phi}_\ell. 
\end{equation}
In addition, we can also solve the constraints \eqref{eq:SU(3)} 
for $\bphi_3$ and obtain
\begin{equation}
 \phi_{a,3} 
 = \ve_{a b c} \overline{\phi}_{b,1} \overline{\phi}_{c,2} .
\end{equation}
This set of equations enable us to eliminate $a_3$ in the theory 
since it can be expressed by the use of $a_1$ and $a_2$ as follows:
\begin{equation}
 \begin{split}
   a_3 
  &= \im \ve_{abc} \phi_{b,1} \phi_{c,2}
  \diff (\ve_{ab'c'} \overline{\phi}_{b',1} \overline{\phi}_{c',2}) 
  \\
  &= - \im  
  \left(
  \overline{\bphi}_{1} \cdot \diff \bphi_{1} 
  + \overline{\bphi}_{2} \cdot \diff \bphi_{2} 
  \right)
  \\
  &= - a_1 - a_2 ,
 \end{split}
\end{equation}
where we used constraints on $\bphi_1$ and $\bphi_2$ to proceed to
the second line.
Therefore, the sum of topological charges is always zero in
the path-integral formula of the partition function.
In other words, the partition function depends only on the difference 
between two theta angles, $\theta_{ij}=\theta_i-\theta_j$.
We can thus take $\theta_3=0$ without loss of generality, 
and consider the phase structure in the $(\theta_1,\theta_2)$-plane.

In order to elucidate the phase structure of 
the $2$-dimensional $SU(3)/U(1)^2$ nonlinear sigma model, 
we need to pay attention to three basic properties of the model: 
global symmetries, 't~Hooft anomaly together with global inconsistency, 
and asymptotic freedom. 
We here summarize these properties.

\medskip
\noindent
\textbf{Global symmetries}: 
The model enjoys four global symmetries; 
$SU(3)/\mathbb{Z}_3$ flavor symmetry, time reversal symmetry $\calT$, 
$\mathbb{Z}_3$ permutation symmetry, and charge conjugation symmetry $\calC$. 
The last two become symmetries only for a class of special thetas and 
for the other class of special thetas, respectively. 
We will briefly summarize these symmetries in order 
(see Ref.~\cite{Tanizaki:2018xto} in more detail). 
\begin{enumerate}
 \item \textbf{$SU(3)/\mathbb{Z}_3$ flavor symmetry}:
       Taking account of the constraints \eqref{eq:SU(3)}, 
       $2$d $SU(3)/U(1)^2$ sigma model is invariant under 
       \begin{equation}
	\bphi_\ell \mapsto U \bphi_\ell \with U \in SU(3).
       \end{equation}
       Furthermore, since its center $\mathbb{Z}_3 \subset SU(3)$---e.g. 
       $U = \rme^{\frac{2\pi i}{3}} \bm{1}_3$---belongs to a part of 
       $U(1)$ gauge symmetries, the correct flavor symmetry is 
       identified as $PSU(3) = SU(3)/\mathbb{Z}_3$.
 \item \textbf{Time reversal symmetry $\calT$}:
       The action \eqref{eq:Action} is shown to be invariant under 
       the time reversal transformation given by 
       \begin{equation}
	\calT: 
	 \begin{cases}
	  \bphi_\ell (x,t) \mapsto \ol{\bphi}_\ell (x,-t),
	  \\
	  a_{\ell, 0} (x,t) \mapsto a_{\ell, 0} (x,-t),
	  \\
	  a_{\ell, 1} (x,t) \mapsto - a_{\ell, 1} (x,-t).
	 \end{cases}
       \end{equation}
       Note that the constraints \eqref{eq:SU(3)} also remain invariant under
       this transformation. 
 \item \textbf{$\mathbb{Z}_3$ permutation symmetry}: 
      We define the $\mathbb{Z}_3$ permutation by 
       \begin{equation}
	\mathbb{Z}_3~\textrm{permutation}:
	 \begin{cases}
	 \bphi_\ell \mapsto \bphi_{\ell + 1} , 
	 \\
	 a_\ell \mapsto a_{\ell + 1},
	 \end{cases}
	 \label{eq:Z3perm}
       \end{equation}
       where we identify the label $\ell$ mod $3$.
	Under this transformation, the action changes as 
	\be
		\Delta S=\im\sum_{\ell=1}^{3}{\theta_{\ell-1}-\theta_{\ell}\over 2\pi}\int \diff a_\ell. 
	\ee
       At the special point in the $(\theta_1,\theta_2)$-plane given by 
       \begin{equation}
			\theta_\ell={2\pi p\over 3}\ell\; \bmod 2\pi,
       \end{equation}
       with $p\in\mathbb{Z}$, the model further enjoys this additional symmetry because $a_1+a_2+a_3=0$ and $\Delta S=0 \bmod 2\pi\im$.   
 \item \textbf{Charge conjugation symmetry $\calC$}: 
       The model also possesses charge conjugation symmetries acting as
       \begin{equation}
	\calC_k: 
	 \begin{cases}
	  \bphi_\ell \mapsto - \ol{\bphi}_{-\ell-k}, 
	  \\
	  a_\ell \mapsto - a_{-\ell-k},
	 \end{cases}
	 (k=1,2,3)
       \end{equation}
       at special points in $(\theta_1,\theta_2)$-plane given by 
       \begin{equation}
	\begin{cases}
	 \calC_1\mathrm{\mathchar`-invariant~points}: 
	 \theta_2 = 2 \theta_1 \hspace{3pt} \mod 2\pi,
	 \\
	 \calC_2\mathrm{\mathchar`-invariant~points}: 
	 \theta_1 = 2 \theta_2  \hspace{3pt} \mod 2\pi,
	 \\
	 \calC_3\mathrm{\mathchar`-invariant~points}: 
	 \theta_1 = - \theta_2 \mod 2\pi.
	 \\
	\end{cases}
	\label{eq:Chargeinv}
       \end{equation}
       Note that all $\calC_k$ becomes symmetry at the $\mathbb{Z}_3$-invariant 
       points.
\end{enumerate}

\medskip
\noindent
\textbf{'t~Hooft anomaly and global inconsistency:} 
Among the above symmetries, there are mixed 't~Hooft anomaly and global 
inconsistency which constrain the possible phase diagram.
In Ref.~\cite{Tanizaki:2018xto}, one of the authors (Y.T.) provided 
intensive study on this constraint, and we here summarize the result.

\begin{enumerate}
 \item \textbf{$SU(3)/\mathbb{Z}_3$-$\mathbb{Z}_3$ anomaly}: 
       As mentioned above, although the action \eqref{eq:Action} enjoys 
       $\mathbb{Z}_3$ permutation symmetry at $\mathbb{Z}_3$-invariant 
       points \eqref{eq:Z3perm}, after gauging $SU(3)/\mathbb{Z}_3$ symmetry,
       the partition function is no longer invariant
       at some $\mathbb{Z}_3$-invariant points. 
	Background gauge fields for $SU(3)/\mathbb{Z}_3$ symmetry consist of~\cite{Aharony:2013hda, Kapustin:2014gua,  Gaiotto:2014kfa, Tachikawa:2014mna} 
	\begin{itemize}
		\item $A$: $SU(3)$ one-form gauge field, and 
		\item $B$: $\mathbb{Z}_3$ two-form gauge field. 
	\end{itemize}
In practical computations~\cite{Tanizaki:2018xto}, we realize the $\mathbb{Z}_3$ two-form gauge field as a pair of $U(1)$ two-form and one-form gauge fields $(B,C)$, satisfying $3B=\diff C$, and embed the $SU(3)$ gauge field to the $U(3)$ gauge field as $\widetilde{A}=A+{1\over 3}C$. By postulating the $U(1)$ one-form gauge invariance, the $U(3)$ bundle can be regarded as $SU(3)/\mathbb{Z}_3$ bundle. 
       The $\mathbb{Z}_3$ permutation symmetry at $\theta_\ell=2\pi p\ell/3$ is anomalously broken as 
       \begin{equation}
	(\mathbb{Z}_3)_{\mathrm{permutation}}:Z[(A,B)] \mapsto
	 Z[(A,B)] \exp \left( - \im p\int_{M_2} B \right) 
	 \at \theta_\ell={2\pi p\over 3}\ell,
	\label{eq:mixed_anomaly_Z3}
       \end{equation}
       where $Z[(A,B)]$ denotes the partition function in the presence of the 
       background $SU(3)/\mathbb{Z}_3$ gauge field.
       This is the mixed 't~Hooft anomaly 
       between $SU(3)/\mathbb{Z}_3$ and $\mathbb{Z}_3$-permutation for $p\not=3$ mod $3$.
       Then, the anomaly matching argument enables us to exclude 
       the trivially gapped ground state. 
       Further taking into account the Coleman-Mermin-Wagner theorem~\cite{Coleman:1973ci, mermin1966absence} and the fact 
       that the topological order is ruled out in $(1+1)$-dimension \cite{PhysRevB.83.035107},
       the possible scenarios are given as follows~\cite{Lajko:2017wif, Tanizaki:2018xto}:
       \begin{itemize}
	\item The ground state spontaneously breaks $\mathbb{Z}_3$ symmetries.
	\item The ground state shows conformal behavior, especially given by $SU(3)_1$ Wess-Zumino-Witten model~\cite{Lajko:2017wif, Tanizaki:2018xto, Ohmori:2018qza}. 
       \end{itemize}
 \item \textbf{$SU(3)/\mathbb{Z}_3$-$\calC$ global inconsistency}: 
       When we can cancel the difference of the partition function
       arising from the gauge transformation by adding local counter terms, 
       it means that there is no 't~Hooft anomaly.
       Nevertheless, there is a situation in which we cannot cover 
       all the coupling constant space by the single local counter terms.
       Then, there is locally no t'~Hooft anomaly, but there is global 
       inconsistency, which also gives restrictions on the phase structure~\cite{Gaiotto:2017yup, Tanizaki:2017bam,Kikuchi:2017pcp}.
       For example, let us consider $\calC_3$-invariant lines 
       $\theta_1 = - \theta_2 + 2\pi k ~(k \in \mathbb{Z})$.
       The $\calC_3$-charge conjugation then induces 
       the change of the partition function given by
       \begin{equation}
	\calC_3: Z[(A,B)] \mapsto 
	 Z[(A,B)] \exp \left( - 2\im k \int_{M_2} B\right).
       \end{equation}
       This difference, however, does not means the 't~Hooft anomaly.
       Indeed, the modified partition function $Z_n[(A,B)]$ 
       with the gauge-invariant local counter terms transforms as 
       \begin{equation}
	Z_n [(A,B)] \equiv Z[(A,B)] \exp \left(\im n \int_{M_2} B \right)
	 \mapsto  
	 Z_n [(A,B)] \exp \left( - 2\im (n+k) \int_{M_2} B \right),
\label{eq:Global_Inconsistency_2d}
       \end{equation}
       and thus, $Z_n [(A,B)]$ remains invariant if $n= 2k$ mod $3$.
       Therefore, there is no 't~Hooft anomaly between $SU(3)/\mathbb{Z}_3$ 
       and $\calC_3$ for a certain fixed $k \in \mathbb{Z}$.
       Nevertheless, when we change $k$ to another value 
       $k' \in \mathbb{Z}$---e.g. $k=0$ to $k'= 1$---we immediately see that 
       another local counter term is necessary to remove the difference. 
       This means we now have $SU(3)/\mathbb{Z}_3$-$\calC_3$ 
       global inconsistency between two lines $(\ell_1,\ell_2)$ defined by 
       $\ell_1: \theta_1 = -\theta_2$ and $\ell_2: \theta_1 = -\theta_2+ 2\pi$.
       Then, we can apply the conjectured matching condition \cite{Tanizaki:2017bam,Kikuchi:2017pcp},
       which states two possibilities:
       \begin{itemize}
	\item The ground state on $\ell_1$ or $\ell_2$ is not trivially gapped.
	\item The ground states on both $\ell_1$ and $\ell_2$ are trivial, 
	      but they are separated by quantum phase transitions, and thus, 
	      belong to different symmetry-protected topological (SPT) phases protected by $SU(3)/\mathbb{Z}_3$. 
       \end{itemize}
       We come to the same conclusion for the $(\calC_1,\calC_2)$-charge 
       conjugation by simply replacing $\theta_2 = 2\theta_1 + 2\pi k$ and 
       $\theta_1 = 2\theta_2 + 2\pi k$ with different $k \in \mathbb{Z}$, 
       and the possible phase structure is constrained~\cite{Tanizaki:2018xto}.
\end{enumerate}

\medskip
\noindent
\textbf{Asymptotic freedom:} 
The symmetry and anomaly argument restricts the possible phase diagram, 
but does not answer which possibility is realized. 
To elucidate the phase structure in details, it is desirable to apply the semiclassical 
analysis to the present theory. 
Nevertheless, there is a troublesome general property for 
$2d$ nonlinear sigma models, that is, the asymptotic freedom.
In fact, the renormalization group (RG) analysis 
in Ref.~\cite{Lajko:2017wif, Ohmori:2018qza}
gives the $\beta$ functions for $g$ and $\lambda$ as 
\begin{equation}
 \beta_g (g,\lambda) = - \frac{5g^2}{4\pi}, 
  \quad
  \beta_\lambda (g,\lambda) = \frac{3g\lambda}{2\pi}
  \with
  \beta_x \equiv \frac{\diff x}{\diff \log\mu},
  \label{eq:RG}
\end{equation}
where $\mu$ is the renormalization scale\footnote{The RG flows obtained in \cite{Lajko:2017wif} and \cite{Ohmori:2018qza} are not consistent for $\lambda$ term while they obtain the same equation for $g$, and we here quote the result from \cite{Ohmori:2018qza} in (\ref{eq:RG}).}.  
Note that $\lambda=0$ is the fixed point of this RG flow\footnote{Although the beta function $\beta_\lambda$ is positive, the \textit{physical} coupling constant is given by $\widetilde{\lambda}=g^{3/2} \lambda$ instead of $\lambda$ itself, as discussed in \cite{Lajko:2017wif, Ohmori:2018qza}, and then $\widetilde{\lambda}$ is the relevant perturbation when we use the result of \cite{Ohmori:2018qza}. Therefore, $\lambda=0$ is the unstable fixed point, so we may need another justification for our setting. As we shall discuss in Sec.~\ref{sec:beyond_DIGA}, the effect of $\lambda$ to the ground-state energy is always subdominant under the $\mathbb{Z}_3$-twisted compactification, which justifies to set $\lambda=0$ in our analysis. }, 
and we will restrict our attention to this case in the following. 
Since $\beta_g (g,\lambda)$ is negative, the model yields 
the asymptotic freedom:
\be
g(\mu)=\left({5\over 4\pi}\ln\left({\mu\over \Lambda}\right)\right)^{-1}. 
\label{eq:RG_running}
\ee
Here, $\Lambda$ is the infrared dynamical scale introduced by the dimensional transmutation. 
This causes a serious problem for the semiclassical analysis 
since it may give a unreliable result even qualitatively~\cite{Witten:1978bc, Affleck:1979gy}.

\subsection{Persistent 't~Hooft anomaly and adiabatic continuation}
\label{sec:Twisted}
There must exist various phase transitions in the phase diagram of 
$2$-dimensional $SU(3)/U(1)^2$ nonlinear sigma model
 in the $(\theta_1,\theta_2)$-plane, due to its rich symmetry structure --- 't~Hooft anomaly 
and global inconsistency. 
However, the asymptotic freedom, or the resulting strong coupling nature 
in the infrared regime, hampers our effort to analytically 
clarify the phase structure of the system.
To circumvent the difficulty caused by the infrared strong dynamics, 
one can compactify the theory with sufficiently small circumference $L \ll \Lambda^{-1}$
to perform the semiclassical analysis.
However, a na\"ive compactification procedure often breaks the 
original symmetry (and 't~Hooft anomaly) structure of the uncompactified theory,
and the phase transition occurs as the compactified size $L$ is varied.
This again prevents us from obtaining a reliable conclusion for the ground state 
of the system.

Much progress has been recently made on semiclassical analysis to obtain reliable results of nonperturbative dynamics. 
The vital point is to employ a proper twisted boundary conditions
associated with the compactification procedure~\cite{Dunne:2012ae, Dunne:2012zk,  Cherman:2013yfa, Misumi:2014jua, Misumi:2014bsa, Sulejmanpasic:2016llc}, and it is conjectured that the phase structure of compactified theory is adiabatically connected to that of uncompactified theory.
It has been also shown that one can systematically construct the compactified 
theory equipped with the same 't~Hooft anomaly as the original theory
by employing the twisted boundary condition~\cite{Tanizaki:2017qhf}, which is consistent with the above conjecture on adiabatic continuity.

In accordance with these developments, let us introduce 
the twisted $SU(3)/U(1)^2$ sigma model~\cite{Tanizaki:2018xto}, 
in which both of the $SU(3)/\mathbb{Z}_3$-$\mathbb{Z}_3$ mixed anomaly and 
$SU(3)/\mathbb{Z}_3$-$\calC$ global inconsistency survive.
We compactify the base manifold as $M_2 = M_1 \times S^1$ with 
circumference $L$,
and introduce the ${\mathbb Z}_{3}$-twisted boundary condition given by
\begin{equation}
 \begin{cases}
 \Phi  (x,t+L) = C \Phi (x,t) , \\
  a_\ell (x,t+L) = a_\ell (x,t),
 \end{cases}
  \with 
  C = \mathrm{diag} \big( 1, \rme^{{2\pi\im/3}}, \rme^{4\pi\im/3} \big).
  \label{eq:TwistedBC}
\end{equation}
We can equivalently describe the twisted theory in terms of 
the field $\tilphi$ satisfying the periodic boundary condition.
For that purpose, we define the untwisted field $\tilbphi_\ell$ as
\begin{equation}
 \tilphi_{f,\ell} (x,t) \equiv 
  \rme^{-2\pi \im ft/3L} \phi_{f,\ell} (x,t) 
  \quad (f=1,2,3),
\end{equation}
and replace the covariant time derivative as
\begin{equation}
 |D_t \phi_{f,\ell}|^2
  = \left| 
     \left( \partial_t + \im a_{\ell, 0} + \frac{2\pi \im f}{3L} \right)
     \tilphi_{f,\ell} \right|^2.
\end{equation}
We can see the equivalence between the ${\mathbb Z}_{3}$-twisted boundary condition and the background flavor-dependent 
$SU(3)$ holonomy along the compactified direction. 

Let us then look into symmetries of the twisted theory. 
Most of the symmetries for the original theory are not affected by 
the twisted boundary condition because they do not depend on the flavor.
Only the original flavor $SU(3)/\mathbb{Z}_3$ symmetry 
is broken down to its maximal Abelian 
subgroup $[U(1)\times U(1)]/\mathbb{Z}_3$ due to the twisted boundary condition,
or the introduction of the background $SU(3)$ holonomy.
We however have another intertwined $\mathbb{Z}_3$-symmetry
coming from the twisted nature of the system.
To see this, it is crucial to note 
$\mathbb{Z}_3 \subset SU(3)$, generated by the shift matrix
\begin{equation}
 \Phi \mapsto S \Phi \with
  S \equiv 
  \begin{pmatrix}
   0 & 1 & 0 \\
   0 & 0 & 1 \\
   1 & 0 & 0
  \end{pmatrix} ,
\label{eq:Z3_shift_phi}
\end{equation}
is \textit{not} symmetry of the twisted theory since it generates the change for 
the kinetic term due to the flavor-dependent background holonomy:
\begin{equation}
 \sum_f 
  \left| 
   \left( \partial_t + \im a_{\ell, 0} + \frac{2\pi \im f}{3L} \right)
   \tilphi_{f,\ell} \right|^2
  \mapsto 
  \sum_f 
 \left|
  \left( \partial_t + \im a_{\ell, 0} + \frac{2\pi \im (f-1)}{3L} \right)
  \tilphi_{f,\ell} \right|^2.
\end{equation}
Nevertheless, we easily see that the difference generated by the shift 
can be absorbed into the $\mathbb{Z}_3$ one-form transformation given by
\begin{equation}
 a_{\ell,0} \mapsto a_{\ell,0} + \frac{2\pi \im}{3L}.
  \label{eq:Z3a}
\end{equation}
Therefore, the system possesses intertwined $\mathbb{Z}_3$ symmetry defined by (\ref{eq:Z3_shift_phi}) and (\ref{eq:Z3a}),
which we will call $(\mathbb{Z}_3)_{\mathrm{shift}}$ symmetry.

By gauging the intertwined $(\mathbb{Z}_3)_{\mathrm{shift}}$ symmetry, 
we can show that the twisted theory indeed shares the structure regarding 't~Hooft anomaly and global inconsistency with the original uncompactified theory.
The crucial point is that the background $\mathbb{Z}_3$ one-form gauge field 
$B^{(1)}$ for the $(\mathbb{Z}_3)_{\mathrm{shift}}$-symmetry is 
directly related to the two-form gauge field $B$ for the flavor 
$SU(3)/\mathbb{Z}_3$-symmetry in the following simple manner:
\begin{equation}
 B = B^{(1)} \wedge L^{-1} \diff t .
\end{equation}
Using (\ref{eq:mixed_anomaly_Z3}), under the $\mathbb{Z}_3$-permutation given in 
eq.~\eqref{eq:Z3perm}, the twisted partition function is shown to transform as
\begin{equation}
 Z_{M_1 \times S^1} [B^{(1)}] 
  \mapsto 
 Z_{M_1 \times S^1} [B^{(1)}] 
 \exp \left( - \im p \int_{M_1} B^{(1)} \right)
 \quad
 \mathrm{at} 
 \quad
\theta_\ell={2\pi p\over 3}\ell.
\end{equation}
This is the  
$(\mathbb{Z}_3)_{\mathrm{shift}}$-$(\mathbb{Z}_3)_{\mathrm{permutation}}$
anomaly, which coincides with the original $SU(3)/\mathbb{Z}_3$-$\mathbb{Z}_3$ 
anomaly.
Similarly, we see the same $(\mathbb{Z}_3)_{\mathrm{shift}}$-$\calC$ 
global inconsistency appears by a simple substitution as
\begin{equation}
 \int_{M_1 \times S^1} B = \int_{M_1} B^{(1)}.
\end{equation}
Therefore, the twisted theory possesses the completely same 't~Hooft 
anomaly and global inconsistency, which again restricts 
the possible phase diagram in $(\theta_1,\theta_2)$-plane.

%---------------------------------------------------------------------------------------
\section{Fractional instantons and Phase diagram}
\label{sec:PhaseDiagram}
In this section, we perform the semiclassical analysis on the twisted 
$2$-dimensional $SU(3)/U(1)^2$ sigma model with $\lambda=0$ and sufficiently small
$L$, and clarify its vacuum phase diagram in the $(\theta_1,\theta_2)$-plane.
In Sec.~\ref{sec:FracInst}, we first construct the fractional 
Bogomol'nyi-Prasad-Sommerfield (BPS) instantons connecting six classical vacua.
In Sec.~\ref{sec:DIGA}, with the help of the dilute instanton gas approximation 
(DIGA), we derive the $\theta_\ell$-dependence of the ground-state energy. 
We there clarify the phase diagram and see that the result of this explicit computation is consistent with the constraint coming out of anomaly and global inconsistency matching.
After demonstrating the accidental enlarged symmetry without $\lambda$-term 
in Sec.~\ref{sec:beyond_DIGA}, we explicitly see how 't~Hooft anomaly and global inconsistency are realized in the semiclassical regime in Sec.~\ref{sec:semiclassical_anomaly}.

\subsection{Classical vacua and fractional instantons}
\label{sec:FracInst}
\begin{figure}[tb]
 \centering
 \includegraphics[scale=0.19]{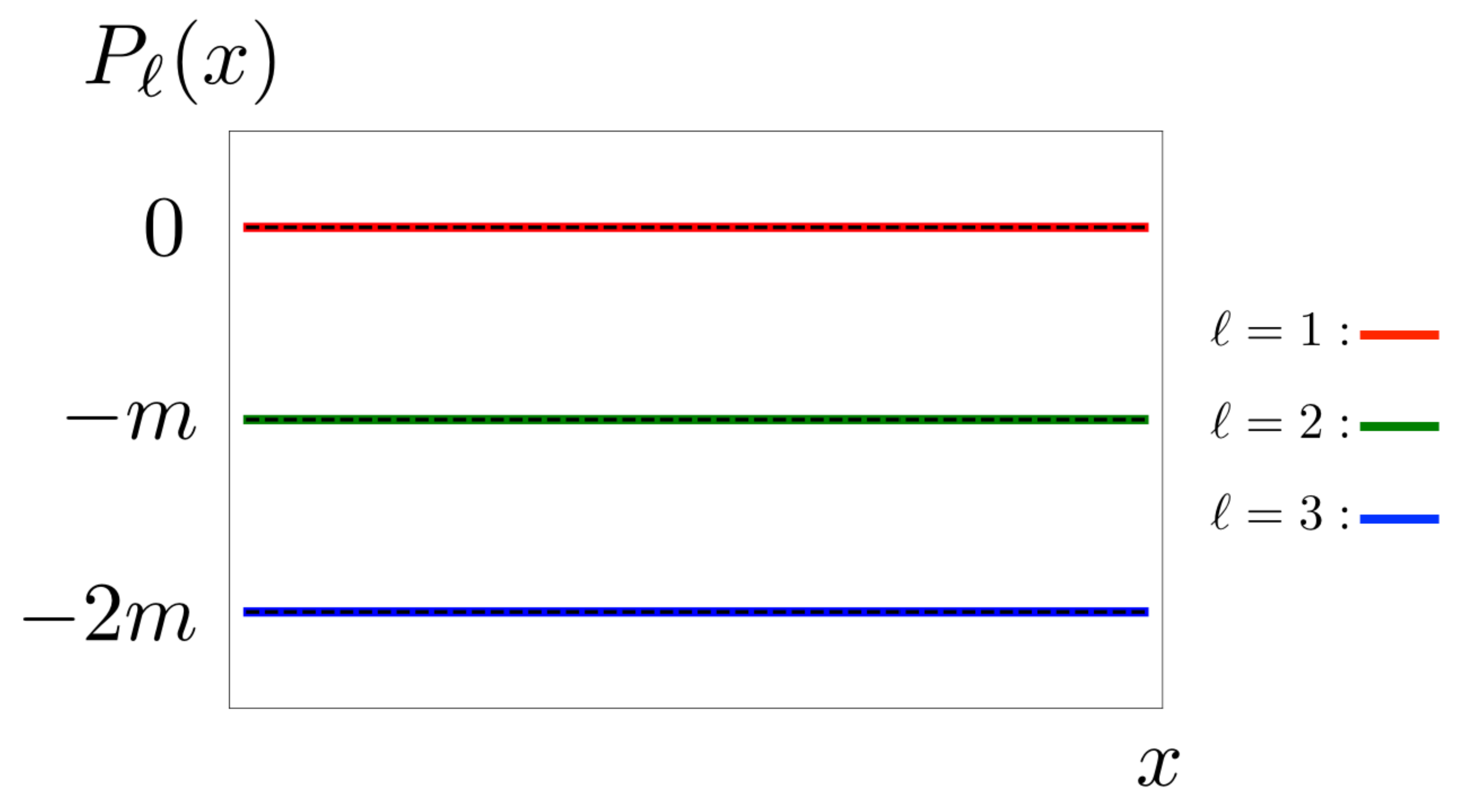}\\
 \caption{ 
 The Polyakov-loop phase $P_\ell (x)$ corresponding to 
 one of the classical vacua $\Phi_1$ in eq.~\eqref{eq:6Vacua}. 
 Red, green, and blue curves correspond to $P_1 (x)$, $P_2(x)$ and $P_3(x)$, 
 respectively.
 }
 \label{fig:vacua}
\end{figure}

In this subsection, 
we first identify the classical vacua and fractional BPS instantons 
connecting them for $\lambda=0$.
In the absence of the $\lambda$-term, 
we can regard the system as the three coupled copies of the $\mathbb{C}P^{2}$ 
nonlinear sigma models due to the constraints \eqref{eq:SU(3)}. 
With the boundary condition (\ref{eq:TwistedBC}), the configurations $\bphi_{\ell}$ with minimal action of $\mathbb{C}P^2$ model are given by 
\be
\bphi_{\ell}=\begin{pmatrix}
1\\ 
0\\
0
\end{pmatrix},\;
\begin{pmatrix}
0\\ 
\rme^{2\pi\im t/3L}\\
0
\end{pmatrix},\;
\begin{pmatrix}
0\\ 
0\\
\rme^{-2\pi\im t/3L}
\end{pmatrix}. 
\ee
Since $\bphi_{\ell}$ ($\ell=1,2,3$) must be orthonormal by the constraint \eqref{eq:SU(3)}, these three $\mathbb{C}P^2$ classical vacua must be assigned to $\bm{\phi}_{\ell}$ by one-to-one correspondence.  
Therefore, the number of classical vacua is $3!=6$, and these six classical vacua $\Phi_a(x,t)$ $(a=1,\ldots,6)$ of the twisted flag sigma model are given by  
\begin{equation}
 \begin{split}
  &\Phi_1=\begin{pmatrix}
	  1 & 0 & 0 \\
	  0 & \rme^{2\pi \im t/3L} & 0 \\
	  0 & 0 & \rme^{-2\pi \im t/3L}
	 \end{pmatrix}, \quad
  \Phi_2=\begin{pmatrix}
	  0 & 1 & 0 \\
	  0 & 0 & \rme^{2\pi \im t/3L} \\
	  \rme^{-2\pi \im t/3L} & 0 & 0
	 \end{pmatrix},\quad\\
	&
  \Phi_3=\begin{pmatrix}
	  0 & 0 & 1 \\
	  \rme^{2\pi \im t/3L} & 0 & 0\\
	  0 & \rme^{-2\pi \im t/3L} & 0
	 \end{pmatrix}, \quad
  \Phi_4=\begin{pmatrix}
	   -1 & 0 & 0 \\
	   0 & 0 & \rme^{2\pi \im t/3L} \\
	   0 & \rme^{-2\pi \im t/3L} & 0
	 \end{pmatrix}, \quad\\
	&
  \Phi_5=\begin{pmatrix}
	  0 & -1 & 0 \\
	  \rme^{2\pi \im t/3L} & 0 & 0 \\
	  0 & 0 & \rme^{-2\pi \im t/3L}
	 \end{pmatrix}, \quad
  \Phi_6=\begin{pmatrix}
	  0 & 0 & -1 \\
	  0 & \rme^{2\pi \im t/3L} & 0\\
	  \rme^{-2\pi \im t/3L} & 0 & 0
	 \end{pmatrix}. 
 \end{split}
 \label{eq:6Vacua}
\end{equation}
We here adopt the expression indicating the ${\mathbb Z}_{3}$-twisted boundary condition explicitly.
Note that the column vectors represent the vacua for each of 
the $\mathbb{C}P^2$ sigma models, and only the restricted combinations appear
due to the the $SU(3)$ constraints \eqref{eq:SU(3)}.
In order to visualize six classical vacua in eq.~\eqref{eq:6Vacua}, 
it is helpful to use the phase of the Polyakov loop (Polyakov-loop phase) along 
the compactified direction, given by
\begin{equation}
 P_\ell (x) \equiv 
  \oint_{S^1} a_{\ell}
  = \im \int_0^L \diff t \ol{\bphi}_\ell \cdot \partial_t \bphi_\ell.
\end{equation}
For example, Fig.~\ref{fig:vacua} shows the configuration of the Polyakov loops
for the vacuum state $\Phi_1$.
Their values are given by a multiple of $m \equiv 2\pi/3$ mod $2\pi$.
Other classical vacua are specified by all possible permutations of 
three colors; red, green, and  blue in Fig.~\ref{fig:vacua}.

Let us then construct the fractional instanton solutions by 
combining a fractional instanton and an fractional anti-instanton
of the $\mathbb{C}P^2$ sigma model~\cite{Fujimori:2016ljw} 
(see appendix~\ref{sec:CPNFracInstanton} for a brief review of $\mathbb{C}P^{N-1}$ fractional instanton solutions). 
For each topological charge $Q_\ell={1\over 2\pi}\int \diff a_\ell$, each kinetic term is bounded from below as 
\be
{1\over 2g}\int |(\diff+\im a_\ell)\bphi_\ell|^2\ge {2\pi\over 2g}|Q_\ell|, 
\ee
and the equality is satisfied for (anti-)BPS solution. 
Because of the constraint $Q_1+Q_2+Q_3=0$, we can designate two topological charges $(Q_1,Q_2)$, and the lower bound of the kinetic term for fixed $(Q_1,Q_2)$ is 
\be
\sum_{\ell=1}^{3}{1\over 2g}\int |(\diff+\im a_\ell)\bphi_\ell|^2\ge {2\pi\over 2g}\left(|Q_1|+|Q_2|+|-Q_1-Q_2|\right). 
\ee
For the finite-action configurations, the topological charges $Q_\ell$ are intimately related to the Polyakov-loop phases at infinities, since they are given by 
\begin{equation}
 Q_\ell 
  \equiv \frac{1}{2\pi} \int \diff a_\ell 
  = - \frac{1}{2\pi} \Big( P_\ell (x=\infty) - P_\ell (x=-\infty) \Big).
\end{equation}
This shows that $Q_\ell\in {1\over 3}\mathbb{Z}$, and thus one of the minimal possibility is $(Q_1,Q_2)=(1/3,-1/3)$. 
This configuration can be constructed explicitly, and the Polyakov-loop profiles are shown in Fig.~\ref{fig:Polyakovloop}: 
\begin{figure}[t]
 \centering
 \includegraphics[scale=0.32]{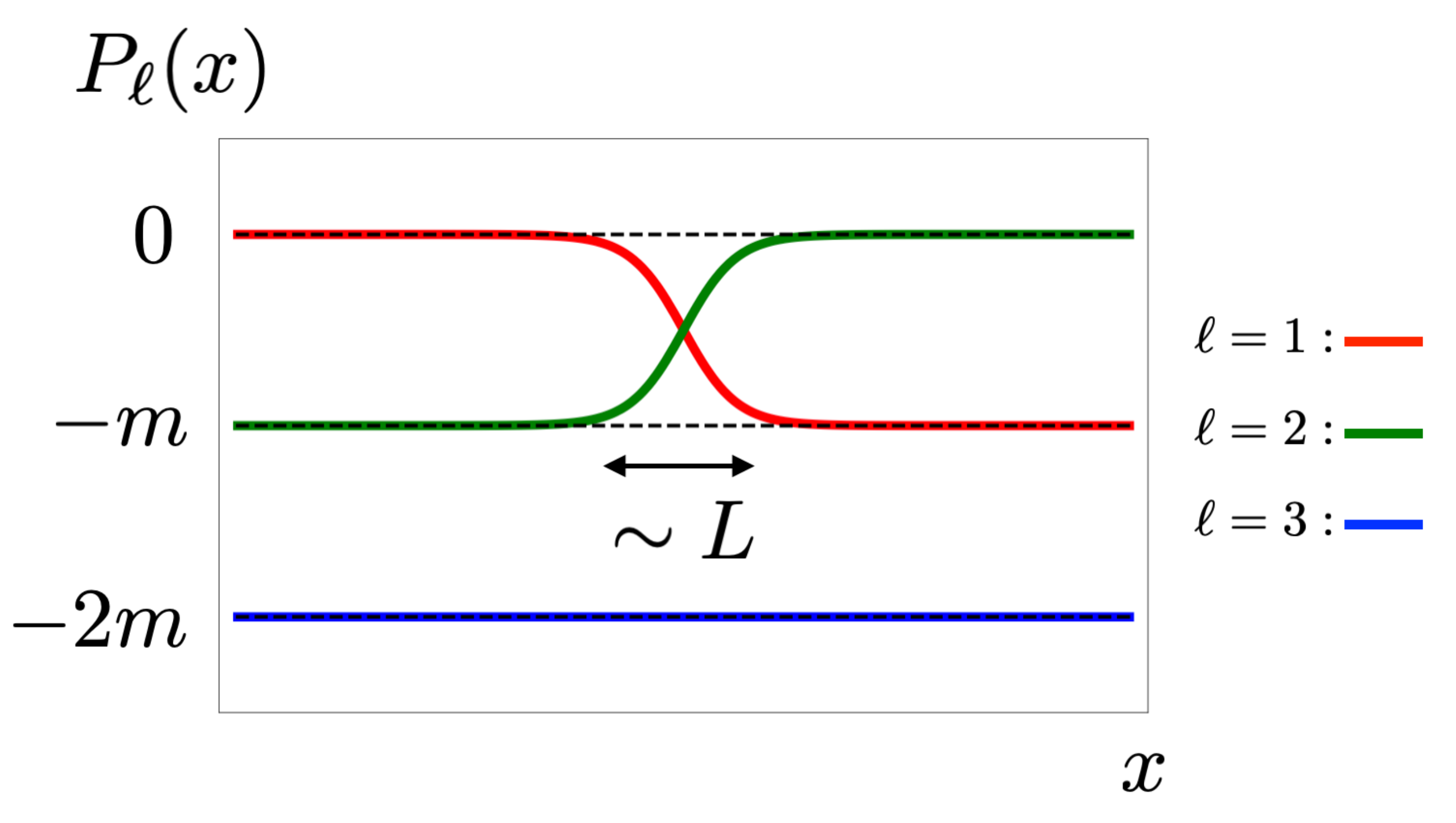}\\
 \caption{
 Spatial dependences of the Polyakov-loop phase for 
 the fractional instanton~\eqref{eq:FracInst}.
 }
 \label{fig:Polyakovloop}
\end{figure}%
Introducing the dimensionless variable $z = {2\pi\over 3L}(x + \im t)$, 
we can construct one solution as
\begin{equation}
 \bphi_1^{\calI} = {1\over \sqrt{1+|\rme^{z-z_0}|^2}}
 \begin{pmatrix}
   1 \\
  \rme^{z-z_0} \\
   0
 \end{pmatrix},\,
 \bphi_2^{\calI} = {1\over \sqrt{1+|\rme^{z_0-z}|^2}}
 \begin{pmatrix}
  - 1 \\
  \rme^{\overline{z}_{0}-\overline{z}} \\
   0 
 \end{pmatrix},\,
 \bphi_3^{\calI} = 
 \begin{pmatrix}
     0 \\
     0 \\
      \rme^{{2\pi \im\over 3L} (t_0-t)}
 \end{pmatrix}, 
 \label{eq:FracInst}
\end{equation}
where $z_0={2\pi\over 3L}(x_0+\im t_0)$ is the moduli parameter:
$x_{0}$ is the position of the fractional instanton while ${2\pi\over 3L}t_{0}$ is a phase modulus.
We again adopt the expression with the ${\mathbb Z}_{3}$-twisted boundary condition being explicit.
One can immediately check that $\bphi_1^{\calI}~(\bphi_2^\calI)$ satisfies the (anti-)BPS instanton equations (see appendix~\ref{sec:CPNFracInstanton}).
Also the constraint \eqref{eq:SU(3)} is trivially satisfied, and thus, 
eq.~\eqref{eq:FracInst} actually gives one BPS fractional instanton solution
for the $SU(3)/U(1)^2$ flag sigma model, which connects classical vacua $\Phi_1$ and $\Phi_5$.
The instanton action $S_\calI$ is obtained by summing magnitude of 
the topological charges:
\begin{equation}
 S_\calI = \sum_{a=1}^3 \frac{2\pi}{2g} |Q_a|
  = \frac{2\pi}{3g}.
\end{equation}
This is the leading nonperturbative contribution in the path integral, and we shall take into account them in the DIGA.

Although we only show one fractional instanton solution---a pair of 
the instanton and anti-instanton for $\bphi_1$ and $\bphi_2$---connecting 
the two classical vacua $\Phi_1$ and $\Phi_5$, it is straightforward to 
construct other solutions by considering all possible combination 
for every two pairs of $\bphi_\ell$.
They are given by a pair of the BPS fractional instanton and anti-instanton 
and connect one of the classical vacua $\Phi_{1,2,3}~(\Phi_{4,5,6})$ to 
another one $\Phi_{4,5,6}~(\Phi_{1,2,3})$.
Thus, the total number of the minimal fractional instanton solutions is $18$.
Note that all the basic BPS fractional instantons have the same value of the 
instanton action, but the different combinations of topological charges.
Although the sum of the topological charge in total again vanishes 
for all solutions, these induce the $\theta_\ell$-dependences of the 
ground-state energy as we will see in the next subsection.

\subsection{Dilute fractional-instanton gas approximation}
\label{sec:DIGA}

We here evaluate the ground-state energy, or the partition function 
with the help of the dilute instanton gas approximation (DIGA).
Before applying the DIGA, let us elucidate its validity. 
Setting the renormalization scale $\mu=1/L$ in the RG running (\ref{eq:RG_running}), we obtain
the nonperturbative contribution coming from the fractional instantons as
\begin{equation}
 \rme^{- S_\calI }
 = \exp \left(- \frac{2\pi}{3g (L^{-1})} \right)
 = (L \Lambda)^{5/6}. 
\end{equation}
The energy gap produced
by the nonperturbative contribution can be evaluated as
\begin{equation}
 \Delta E_{\mathrm{np}} \sim L^{-1} (L \Lambda)^{5/6}.
\end{equation}
On the other hand, the perturbative contribution typically leads to
the energy $E_{\mathrm{p}} \sim L^{-1}$.
Therefore, in order to apply the DIGA, the condition
\begin{equation}
 \Delta E_{\mathrm{np}} \ll  E_{\mathrm{p}}  
  \quad \Leftrightarrow \quad
  (L \Lambda)^{5/6} \ll 1,
  \label{eq:DIGAcondition}
\end{equation}
should be satisfied.
This indicates that the DIGA works well if we assume the sufficiently small compactified 
size $L$ satisfying \eqref{eq:DIGAcondition}.
We note that the condition \eqref{eq:DIGAcondition} is equivalent to 
the simple condition $\rme^{- S_\mathcal{I}} \ll 1$.

To obtain the ground-state energy, we compute the transition amplitude, 
\be
\langle\Phi_a| \exp(-\beta H)|\Phi_b\rangle=\int_{\Phi(0)=\Phi_b}^{\Phi(\beta)=\Phi_a} \Diff \Phi(x)\exp\left(-S[\Phi(x)]\right), 
\ee
where $\ket{\Phi_a}$ denotes the classical vacuum states characterized 
by $\Phi_a$ in eq.~\eqref{eq:6Vacua}. 
In the limit $L^{-1}\rme^{-S_\calI}\ll \beta^{-1} \ll L^{-1}$, the transition amplitude is well approximated by one-fractional instanton transitions:  
\begin{equation}
\beta \begin{pmatrix}
  0 & M^\dagger\\
  M & 0
 \end{pmatrix}
 \with 
 M= K \rme^{-S_\calI}
 \begin{pmatrix}
  \rme^{{\im\over 3}(\theta_2-\theta_3)}&\rme^{{\im\over 3}(\theta_1-\theta_2)}& \rme^{{\im\over 3}(\theta_3-\theta_1)}\\
  \rme^{{\im\over 3}(\theta_1-\theta_2)}& \rme^{{\im\over 3}(\theta_3-\theta_1)} & \rme^{{\im\over 3}(\theta_2-\theta_3)}\\
  \rme^{{\im\over 3}(\theta_3-\theta_1)} & \rme^{{\im\over 3}(\theta_2-\theta_3)} & \rme^{{\im\over 3}(\theta_1-\theta_2)}
 \end{pmatrix},
\label{eq:One_Instanton_Amplitude}
\end{equation}
where $\rme^{-S_\calI}$ represents the common nonperturbative contribution 
coming from the fractional instantons, and 
$K$ is a prefactor coming from one-loop determinant.
Note that the phase factors depending on the differences between $\theta_\ell$ 
appear in the matrix $M$.
We here keep all $\theta_\ell$, but we will eventually set $\theta_3 = 0$.
In the DIGA, the transition amplitude with $\beta L^{-1}\gg 1$ is approximated by using the one-instanton matrix element as follows:
\begin{equation}
 \bra{\Phi_a}\exp(-\beta H) \ket{\Phi_b}
  =\exp \left(\beta 
	 \begin{pmatrix}
	  0&M^\dagger\\
	  M &0
	 \end{pmatrix}\right). 
\end{equation}
Using this amptlitude, we can compute the partition function as
\begin{equation}
 \begin{split}
  Z&= \tr \left[\exp\left(\beta
  \begin{pmatrix}
   0&M^\dagger\\
   M &0
  \end{pmatrix} \right)\right]
  \\
  &=\tr\left[\exp(\beta \sqrt{M^\dagger M})\right]+\tr\left[\exp(-\beta \sqrt{M^\dagger M})\right]. 
 \end{split}
\end{equation}

\begin{figure}[tb]
 \centering
  \includegraphics[scale=0.5]{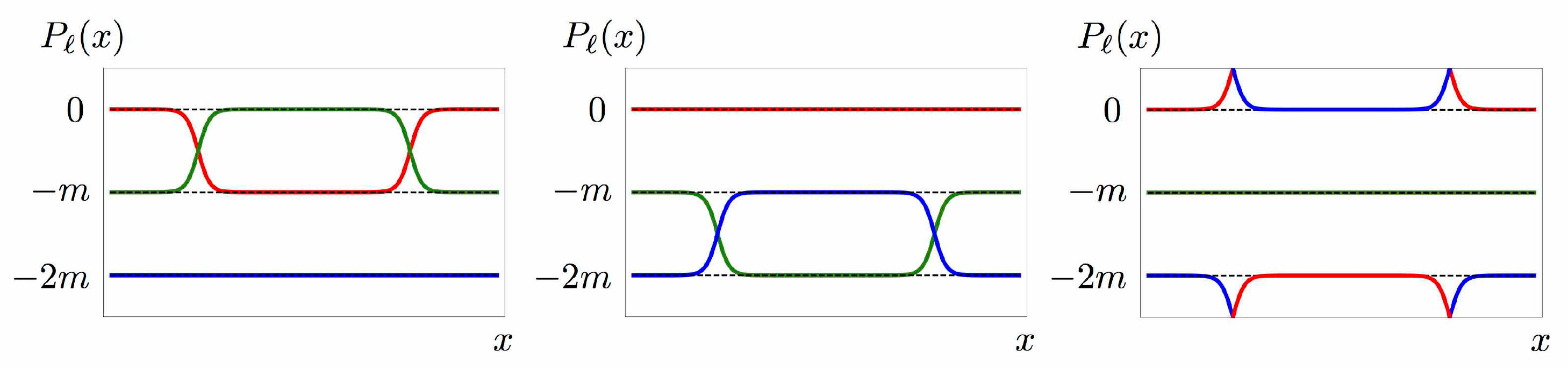}\\
 \caption{
 Polyakov-loop phases for two pairs of the fractional instantons starting 
 from the classical vacuum $\Phi_1$ given in eq.~\eqref{eq:6Vacua}. 
 Red, green, and blue curves correspond to $P_1 (x)$, $P_2(x)$ and $P_3(x)$, 
 respectively.
 }
 \label{fig:2instantons}
\end{figure}

\begin{figure}[tb]
 \centering
 \includegraphics[scale=0.5]{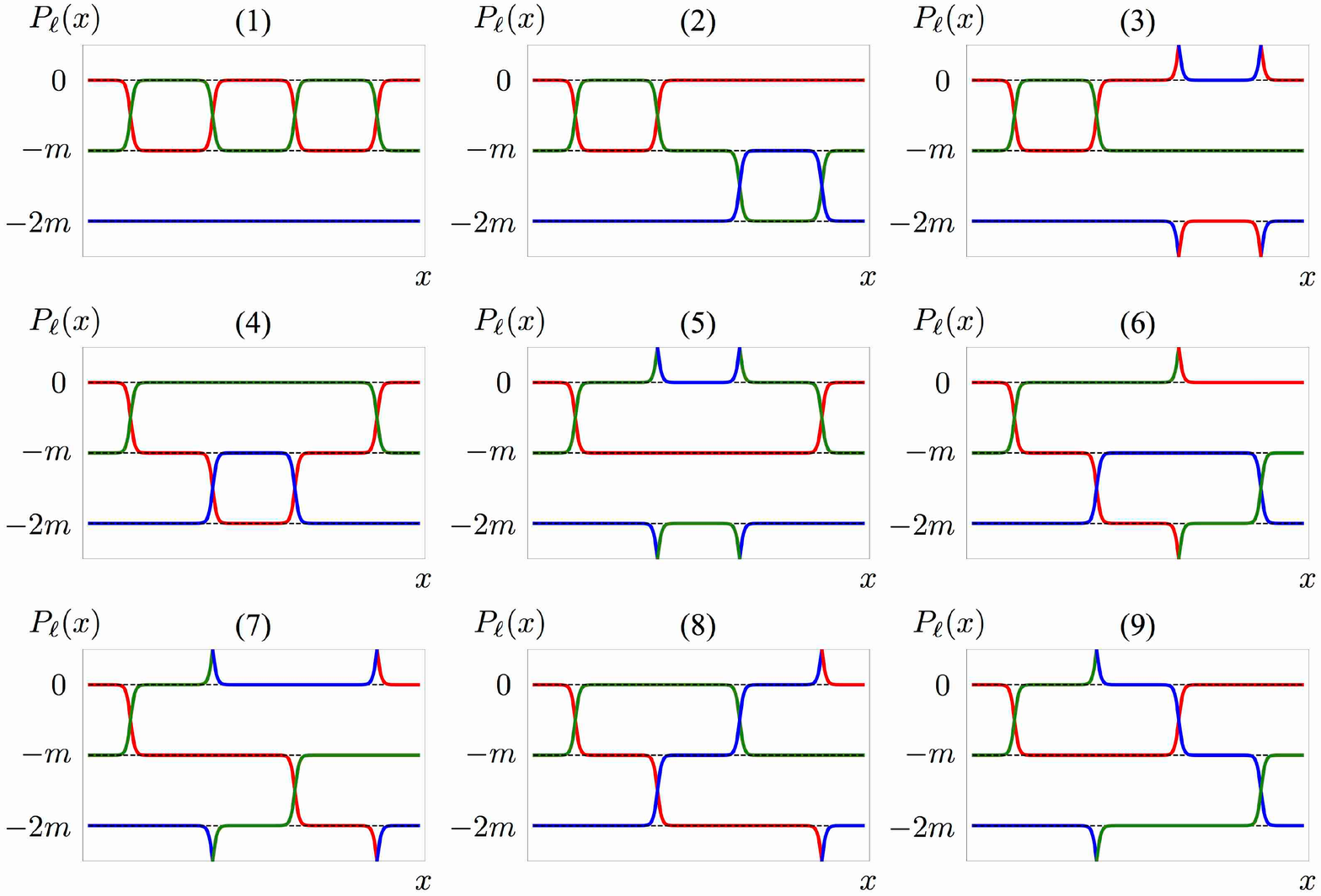}\\
 \caption{
 Polyakov-loop phases  for four pairs of the fractional 
 instantons starting from and ending at the classical vacuum $\Phi_1$ 
 in eq.~\eqref{eq:6Vacua} with the fractional instantons \eqref{eq:FracInst}. 
 At this order, there are $27$ possible configurations, but we here only show $9$ of them by restricting the first jump of fractional instantons as $\Phi_1\to \Phi_5$. 
 }
 \label{fig:4instantons}
\end{figure}

Before evaluating the eigenenergies, we here check the consistency 
between our calculation and the direct DIGA calculation.
To see this, we expand the partition function as 
\begin{equation}
 \begin{split}
  Z 
  &= 6 \sum_{n\geq 0} \frac{\beta^{2n}}{(2n)!} 
  \bra{\Phi_1} (M^\dag M)^n \ket{\Phi_1}
  \\
  &= 6\left( Z_{\mathrm{LO}} + Z_{\mathrm{NLO}} + \calO (e^{- 6 S_\calI }) \right),
 \end{split}
 \label{eq:ComparingZ}
\end{equation}
where we used the consequence originating in $\mathbb{Z}_3$ symmetry  
to obtain the first expression:
\begin{equation}
 \bra{\Phi_1} (M^\dag M)^n \ket{\Phi_1}
  = \bra{\Phi_2} (M^\dag M)^n \ket{\Phi_2}
  = \bra{\Phi_3} (M^\dag M)^n \ket{\Phi_3}.
\end{equation}
Here, $Z_{\mathrm{LO}}$ ($Z_{\mathrm{NLO}}$) denotes the contribution
from two (four) pairs of the fractional instantons starting and ending at a fixed classical vacuum, say, $\Phi_1$. They follows from 
the transition amplitude \eqref{eq:One_Instanton_Amplitude}:
\begin{align}
  Z_{\mathrm{LO}}
  &= {1\over 2!}\left(\beta K e^{- S_\calI}\right)^2 \times 3 , 
 \label{eq:DIGALO}
  \\
  Z_{\mathrm{NLO}}
  &= \frac{1}{4!} \left(\beta K e^{- S_\calI}\right)^4 \times 
  \big[ 15 + 4\cos (\theta_1 - \theta_2) + 4 \cos (\theta_2 - \theta_3 ) 
  + 4 \cos (\theta_3 - \theta_1) \big].
  \label{eq:DIGANLO}
\end{align}
On the other hand, we can count all the leading-order and next-leading-order 
configurations in the DIGA by looking into their profiles of Polyakov loop phases.
Fig.~\ref{fig:2instantons} shows all the leading-order profiles with 
two pairs of the fractional instantons. Fig.~\ref{fig:4instantons} shows 
nine representative ones among the 27 configurations in the next-leading order with four pairs of the fractional instantons, by restricting the first jump of instanton as $\Phi_1\to \Phi_5$ among three possibilities $\Phi_1\to \Phi_{4,5,6}$.
We clearly see there is no nontrivial winding in the leading-order 
configurations while some of next-leading-order ones have (see configurations (6)-(9) in Fig.~\ref{fig:4instantons}). 
These are completely consistent with the $\theta_\ell$-dependence of the partition function
given in eq.~\eqref{eq:DIGALO}-\eqref{eq:DIGANLO}.
This indicates that our calculation is indeed equivalent to the direct 
calculation of the DIGA.

\begin{figure}[t]
 \centering
  \includegraphics[scale=0.5]{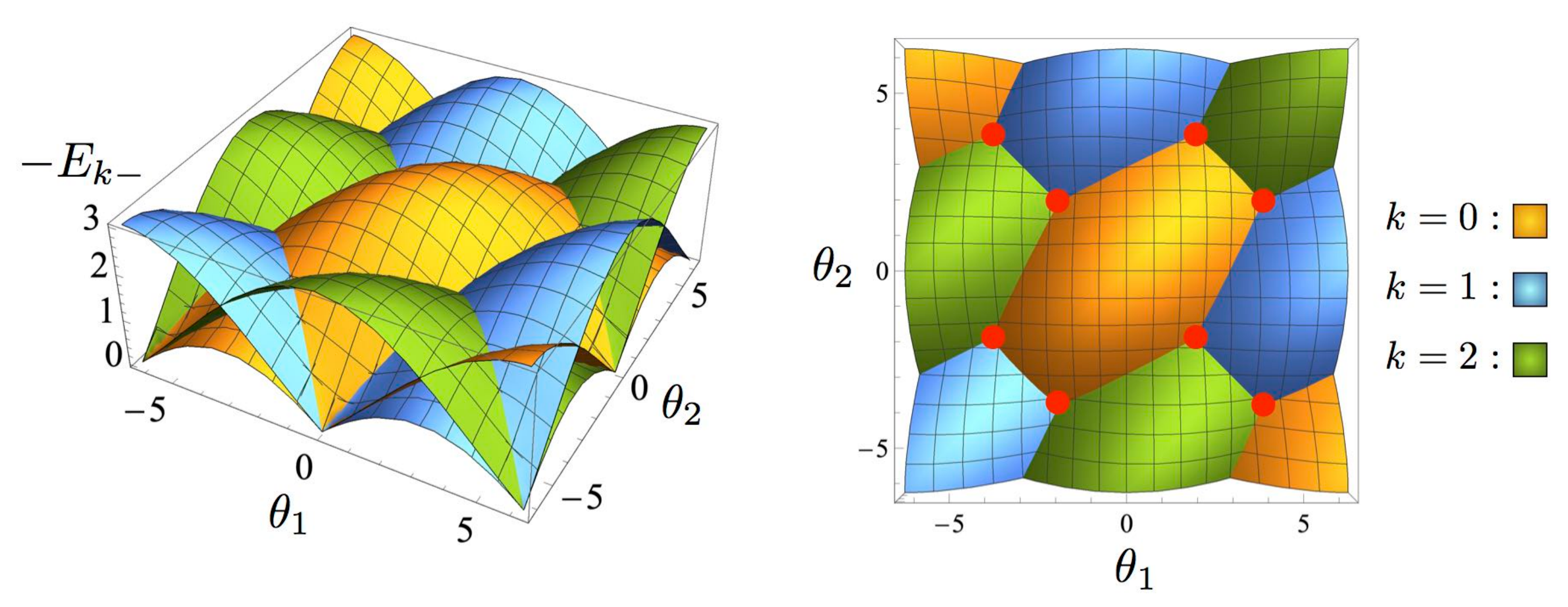}\\
 \caption{
 The ground-state energy of the $2$-dimensional twisted $SU(3)/U(1)^2$ 
 sigma model $(\lambda=0)$ in the $(\theta_1,\theta_2)$-plane.
 The quantum phase-transition lines are located in the places consistent with 
 the anomaly/global inconsistency matching.
 Furthermore, we also see the triple degeneracies of the ground-state energy 
 appear at special points e.g. $(2\pi /3,-2\pi /3)$.
 }
 \label{fig:Phasediag}
\end{figure}

Let us then evaluate the eigenenergies by diagonalizing the matrix $M^\dag M$.
The partition function indicates that the six lowest eigenenergies of the system $E_{k \pm}~(k=0,1,2)$ 
can be obtained by solving the characteristic equation 
\begin{equation}
 0=\det\left[(E_{k\pm})^2 \bm{1}_3-M^\dagger M\right]. 
\end{equation}
Setting $\theta_3=0$, we now obtain these six eigenenergies as
\begin{equation}
 E_{k\pm}(\theta_1,\theta_2)
  =\pm K \rme^{-{S_\mathcal{I}}}
 \Bigl|
	\rme^{{\im\over 3}(\theta_1-\theta_2)}+\rme^{{\im\over 3}(\theta_2+2\pi k)} +\rme^{-{\im\over3}(\theta_1+2\pi k)}
 \Bigr|. 
  \label{eq:Eigenenergy}
\end{equation}
As we shall see in Sec.~\ref{sec:semiclassical_anomaly}, we can confirm that each state is the eigenstate of $(\mathbb{Z}_3)_{\mathrm{shift}}$ symmetry, with the eigenvalue $\rme^{-2\pi\im k/3}$. Indeed, the corresponding energy eigenstates are given by
\begin{equation}
 |E_{k\pm}\rangle
  = \sqrt{f_k^*(\theta_1,\theta_2)}\left(\omega^{k}|\Phi_1\rangle +|\Phi_2\rangle+\omega^{-k}|\Phi_3\rangle\right)
 \pm \sqrt{f_k(\theta_1,\theta_2)}\left(|\Phi_4\rangle +\omega^{k} |\Phi_5\rangle+\omega^{-k} |\Phi_6\rangle\right),
\label{eq:Eigenstate}
\end{equation}
with $\omega=\rme^{2\pi\im/3}$ and $f_k(\theta_1,\theta_2)=\rme^{{\im\over 3}(\theta_1-\theta_2)}+\rme^{{\im\over 3}(\theta_2+2\pi k)} +\rme^{-{\im\over3}(\theta_1+2\pi k)}$. 
The ground-state energy has the three-branch structure:
\begin{equation}
 E_{\mathrm{gs}} (\theta_1,\theta_2) = \min_{k\in\{0,1,2\}} E_{k-}(\theta_1,\theta_2).  
  \label{eq:GSenergy}
\end{equation}
This three-branch structure is shown in Fig.~\ref{fig:Phasediag}

Fig.~\ref{fig:Phasediag} shows the profile of the ground-state energy
and resulting phase diagram in the $(\theta_1,\theta_2)$-plane.
We clearly see that quantum phase transitions take place on 
the special line segments (lines between different color regions 
in the right figure), 
which induce the three different types of the SPT phases.
In addition, at the special points---like $(\theta_1,\theta_2) = (2\pi/3,-2\pi/3)$---shown as red points in the right figure, 
the eigenenergies $E_{k-}~(k=0,1,2)$ take the same value, 
which means the triple degeneracy is realized there. 
These are completely consistent with the conjectured phase diagram 
by the anomaly/global inconsistency matching. This point will be discussed in Sec.~\ref{sec:semiclassical_anomaly} in more detail. 

The obtained phase diagram has the same structure with the ones for the lattice strong-coupling limit~\cite{Lajko:2017wif} and for the 
$SU(3)/U(1)^2$ \textit{linear} sigma model with heavy scalars~\cite{Tanizaki:2018xto}, but we emphasize that this is the first result 
in which the matter field $\bphi_\ell$ can be regarded as would-be 
Nambu-Goldstone bosons.

\subsection{Speculation on effects of $\lambda$-term and higher KK fractional instantons}
\label{sec:beyond_DIGA}

In this paper, we mainly focus on the DIGA of the twisted $SU(3)/U(1)^2$ sigma model, where we turn off the $\lambda$-term. 
In view of properties of the ground states, we have seen that our result reproduces the expectation from anomaly and global inconsistency conditions given in Ref.~\cite{Tanizaki:2018xto}, and it is quite successful. 
We here provide detailed properties of the energy spectra based on 
the symmetry accidentally enlarged with vanishing $\lambda$-term.

As an example, let us set $\theta_1=\theta_2=0$ in (\ref{eq:Eigenenergy}), then the energy eigenvalues are 
\be
E_{0\pm}=\pm 3 K \rme^{-S_{\mathcal{I}}},\; E_{1\pm}=E_{2\pm}=0. 
\ee
This shows the unique ground state $E_{0-}$, but the first excited states are four-fold degenerate. 
Two-fold degeneracies of $E_{1+}=E_{2+}$ and of $E_{1-}=E_{2-}$ are expected from the existence of charge-conjugation symmetry $\mathcal{C}$, but the DIGA realizes additional degeneracy, say, $E_{1+}=E_{1-}$ (or $E_{2+}=E_{2-}$). 
A related fact is that the expression of eigenstates (\ref{eq:Eigenstate}) is singular when $\theta_1=\theta_2=0$, since the coefficient $f_{1,2}(0,0)=0$. 

Let us more clearly show that this additional degeneracy is accidental. At $\theta_1=\theta_2=0$, the full symmetry group of twisted $SU(3)/U(1)^2$ sigma model is 
\be
\Bigl((\mathbb{Z}_3)_{\mathrm{shift}}\times (\mathbb{Z}_3)_{\mathrm{permutation}}\Bigr)\rtimes (\mathbb{Z}_2)_{\mathcal{C}}. 
\ee
Irreducible representations of this group are $1$ or $2$ dimensional representations, and thus four-fold degeneracy must be accidental in this viewpoint. The above four-fold degeneracy, however, is a consequence of our setting $\lambda=0$, because $(\mathbb{Z}_3)_{\mathrm{permutation}}$ is enlarged to the symmetric group $S_3$ when $\theta_1=\theta_2=0$ and $\lambda=0$~\cite{Ohmori:2018qza}: The full symmetry is 
\be
\Bigl((\mathbb{Z}_3)_{\mathrm{shift}}\times S_3\Bigr)\rtimes (\mathbb{Z}_2)_{\mathcal{C}}. 
\ee
This symmetry group has a four-dimensional irreducible representation, which explains the above degeneracy of the first excited state. 

This analysis indicates that we must remove our assumption $\lambda=0$ to solve the degeneracy between $E_{1+}$ and $E_{1-}$. Interestingly, this consequence, at the same time, requires the analysis beyond the lowest Kaluza-Klein (KK) mode approximation. To see this, let us assume that we pick up one KK mode for each $\bphi_\ell$, then $\overline{\bphi_{\ell+1}}\cdot \p_t \bphi_\ell\propto \overline{\bphi_{\ell+1}}\cdot \bphi_\ell=0$ by the orthogonality constraint. Since the $\lambda$-term must contain $\overline{\bphi_{\ell+1}}\cdot \p_t\bphi_\ell$ by its topological nature, it automatically vanishes in this approximation. 

\begin{figure}[tb]
 \centering
 \includegraphics[scale=0.40]{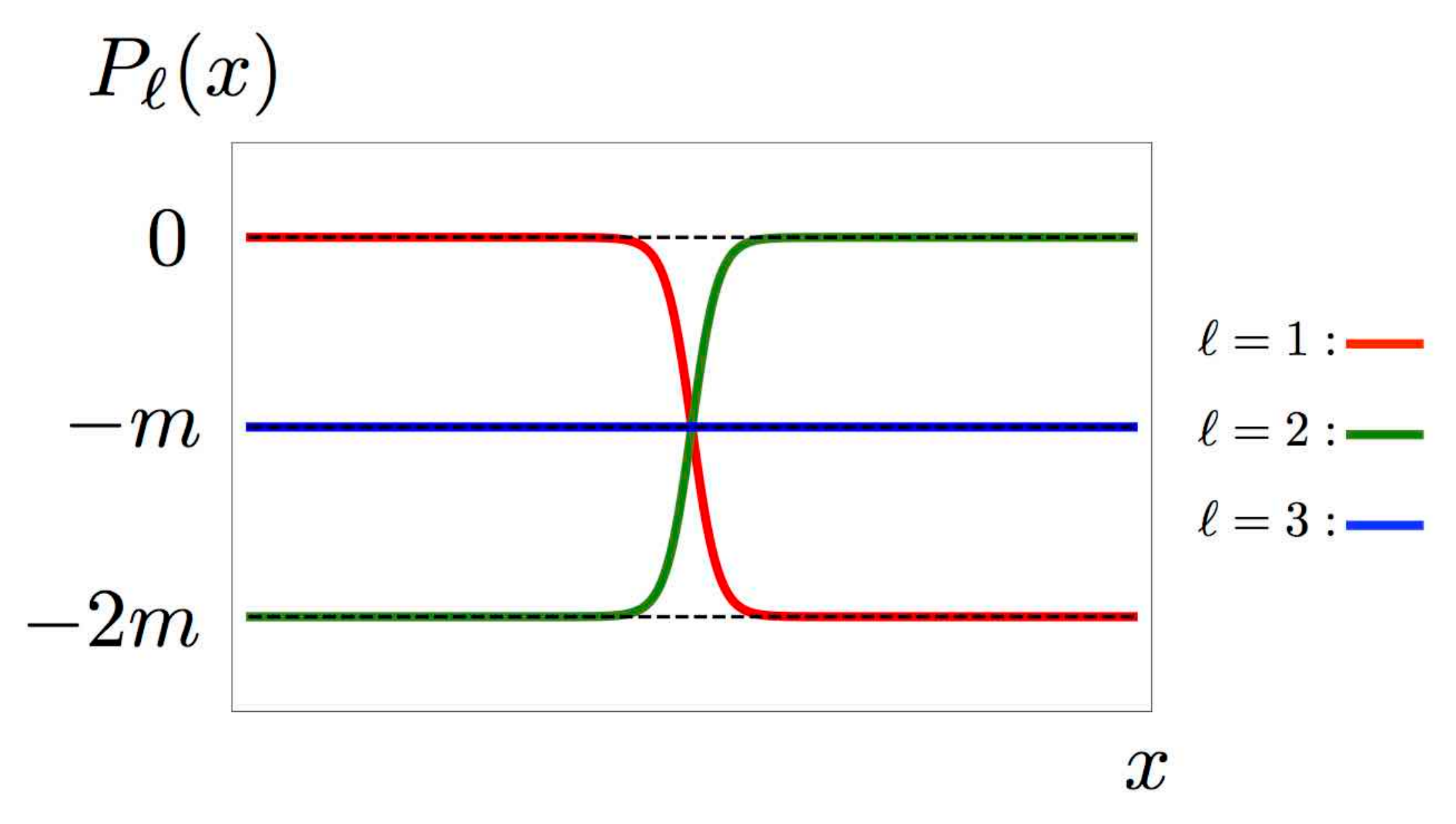}\\
 \caption{ 
 Polyakov loop phases $P_\ell (x)$ for the higher KK fractional instanton 
 \eqref{eq:FracInst_higherKK}.
 }
 \label{fig:higherKKInst}
\end{figure}

Lastly, let us give an example of higher KK fractional instantons, which may play an important role in the study of the first excited state at $\theta_1=\theta_2=0$ with $\lambda\not=0$. 
\begin{eqnarray}
&& \bphi_1^{\calI} = {1\over \sqrt{1+|\rme^{2(z-z_0)}|^2}}
 \begin{pmatrix}
   1 \\
   0 \\
  \rme^{2(z-z_0)}
 \end{pmatrix},\,\quad\quad
 \bphi_2^{\calI} = {1\over \sqrt{1+|\rme^{2(z_0-z)}|^2}}
 \begin{pmatrix}
  1 \\
   0 \\
  -\rme^{2(\overline{z}_{0}-\overline{z})}
 \end{pmatrix},\,\nonumber\\
&& \bphi_3^{\calI} = 
 \begin{pmatrix}
     0 \\
      \rme^{{4\pi \im\over 3L} (t_0-t)}\\
	0
 \end{pmatrix}. 
 \label{eq:FracInst_higherKK}
\end{eqnarray}
Fig.~\ref{fig:higherKKInst} shows the Polyakov phase attached to
the higher KK fractional instantons \eqref{eq:FracInst_higherKK}.
Note that $\bphi_3^{\calI}$ correctly satisfies the ${\mathbb Z}_{3}$-twisted boundary condition since $e^{-{4\pi i \over{3}}} = e^{{2\pi i \over{3}}}$.
The topological charge of this fractional instanton is $(Q_1,Q_2)=(2/3,-2/3)$, and the action is 
\be
S=2S_{\mathcal{I}}={4\pi\over 3g}. 
\ee
This BPS solution cannot be constructed by combining two minimal fractional instantons, and thus we have to go beyond the DIGA to discuss its effect. 
We have argued its possible importance on excited states, but we also would like to point out that such an effect is parametrically smaller than one-instanton contributions for most of the theta angles. 
Especially, its effect on the ground state energies are always subdominant, and thus it does not change the consequence of our analysis on the ground-state properties of the twisted partition function.

\subsection{Semi-classical realization of anomaly and global inconsistency}
\label{sec:semiclassical_anomaly}

In the DIGA, we approximate the Hilbert space and the Hamiltonian as  
\be
\mathcal{H}=\sum_{a=1}^{6}\mathbb{C}|\Phi_a\rangle\simeq \mathbb{C}^6,\quad
H=\begin{pmatrix}
0&M^\dagger\\
M&0
\end{pmatrix}, 
\ee 
and $M$ is given in (\ref{eq:One_Instanton_Amplitude}). 
We can explicitly see how the anomaly and global inconsistency are realized in the energy spectrum (\ref{eq:Eigenenergy}) by constructing the symmetry algebra on $\mathcal{H}$. 
The anomaly matching argument states that we should be able to reproduce the anomaly in Sec.~\ref{sec:Twisted}. 

The $(\mathbb{Z}_3)_{\mathrm{shift}}$  symmetry classically acts on $\Phi_a$ as 
\begin{equation}
 (\mathbb{Z}_3)_{\mathrm{shift}}:
  \begin{cases}
   \Phi_1 \mapsto \Phi_2, \quad 
   \Phi_2 \mapsto \Phi_3, \quad 
   \Phi_3 \mapsto \Phi_1,
   \\
   \Phi_4 \mapsto \Phi_6, \quad 
   \Phi_5 \mapsto \Phi_4, \quad 
   \Phi_6 \mapsto \Phi_5. 
   \\
  \end{cases}
\end{equation}
This motivates us to define the operator on $\mathcal{H}$ as 
\be
U=\begin{pmatrix}
S&0\\
0&S^{-1}
\end{pmatrix},
\with
  S \equiv 
  \begin{pmatrix}
   0 & 1 & 0 \\
   0 & 0 & 1 \\
   1 & 0 & 0
  \end{pmatrix},
\ee
where we again used $\omega = e^{2\pi \im/3}$.
Since $S M S=M$, we find that $U$ generates the $(\mathbb{Z}_3)_{\mathrm{shift}}$ symmetry: $U^3=1$ and 
\be
U^{\dagger} H U=\begin{pmatrix}
0&S^{-1}M^\dagger S^{-1}\\
SMS & 0
\end{pmatrix}=H.
\ee
This acts on the eigenstate $|E_{k\pm}\rangle$ as $U|E_{k\pm}\rangle= \omega^{-k} |E_{k\pm}\rangle$. 

\medskip
\noindent
\textbf{$(\mathbb{Z}_3)_{\mathrm{shift}}$-$(\mathbb{Z}_3)_{\mathrm{permutation}}$ anomaly:} 
At the special point of theta angles, $\theta_\ell={2\pi p\over 3}\ell \bmod 2\pi$, we have an extra symmetry $(\mathbb{Z}_{3})_{\mathrm{permutation}}$. It classically acts on $\Phi_a$ as 
\begin{equation}
 (\mathbb{Z}_3)_{\mathrm{permutation}}:
  \begin{cases}
   \Phi_1 \mapsto \Phi_2, \quad 
   \Phi_2 \mapsto \Phi_3, \quad 
   \Phi_3 \mapsto \Phi_1,
   \\
   \Phi_4 \mapsto \Phi_5, \quad 
   \Phi_5 \mapsto \Phi_6, \quad 
   \Phi_6 \mapsto \Phi_4. 
  \end{cases}
\end{equation}
When acting on the states $|\Phi_a\rangle$, we need to multiply appropriate $\mathbb{Z}_3$ phase in addition to this classical transformation. 
As an example, let us take $(\theta_1,\theta_2)=({2\pi\over 3}p,-{2\pi\over 3}p)$ with $p\in\mathbb{Z}$, then the submatrix $M$ of Hamiltonian $H$ is given by
\be
 M = K e^{-S_\calI}
\begin{pmatrix}
\omega^{-p/3} & \omega^{2p/3} & \omega^{-p/3}\\
\omega^{2p/3} & \omega^{-p/3} & \omega^{-p/3}\\
\omega^{-p/3} & \omega^{-p/3} & \omega^{2p/3}
\end{pmatrix}. 
\ee 
Let us define the $\mathbb{Z}_3$ transformation 
\be
V_p=\begin{pmatrix}
S_p & 0\\
0 & S_{-p}
\end{pmatrix}, 
\with
  S_p \equiv 
  \begin{pmatrix}
   0 & 1 & 0 \\
   0 & 0 & \omega^{-p} \\
   \omega^p & 0 & 0
  \end{pmatrix}, 
\ee
then this is the symmetry at $(\theta_1,\theta_2)=({2\pi\over 3}p,-{2\pi\over 3}p)$ because we can check that 
$V_p^\dagger H V_p = H$ by using $S_{-p}^\dagger M S_{p}=M$. $V_p$ acts on the eigenstates as $V_p |E_{k\pm}\rangle \propto |E_{(k+p)\pm}\rangle$. 

Since $S S_{p} S^{-1}=\omega^{-p} S_p$, $U$ and $V_p$ does not commute for $p\not=0\bmod 3$: 
\be
U V_p U^{-1}=\omega^{-p}V_p. 
\label{eq:Projective}
\ee
Therefore, $(\mathbb{Z}_3)_{\mathrm{shift}} \times (\mathbb{Z}_3)_{\mathrm{permutation}}$ has the projective representation for $p\not=0\bmod 3$, which requires the triple degeneracy of the states. 
The red blobs of Fig.~\ref{fig:Phasediag} indeed shows this degeneracy. The projective phase $\omega^{-p}$ of (\ref{eq:Projective}) is nothing but the consequence of anomaly discussed in Sec.~\ref{sec:Twisted}. 

\medskip
\noindent
\textbf{$(\mathbb{Z}_3)_{\mathrm{shift}}$-$\calC$ global inconsistency:} 
On special lines given in eq.~\eqref{eq:Chargeinv}, 
the system does not have the $(\mathbb{Z}_3)_{\mathrm{permutation}}$ symmetry
in general, but still has the charge-conjugation symmetry.
Let us consider lines $\ell_n: \theta_1 + \theta_2=2\pi n$, then the submatrix $M$ 
takes the following forms:
\begin{equation}
 M_{\ell_n} 
  = K \rme^{-S_\calI}
  \begin{pmatrix}
   \omega^n \rme^{-{\im\over 3} \theta_1} & \omega^{-n}\rme^{{2\im\over 3}\theta_1} & \rme^{-{\im\over 3}\theta_1}\\
   \omega^{-n} \rme^{{2\im\over 3}\theta_1} & \rme^{-{\im\over 3}\theta_1} & \omega^n \rme^{-{\im\over 3}\theta_1}\\
   \rme^{-{\im\over 3}\theta_1} & \omega^n \rme^{-{\im\over 3}\theta_1} & \omega^{-n}\rme^{{2\im\over 3}\theta_1}
  \end{pmatrix},
\end{equation}
On these lines $\ell_n$, the system enjoys 
the $\calC_3$-charge conjugation symmetry. 
With trial-and-error, we find that its representation matrix is 
\be
\mathcal{C}_{\ell_n}=\begin{pmatrix}
C_n&0\\
0&C_{-n}
\end{pmatrix}, 
\with 
C_n=\begin{pmatrix}
0&\omega^n&0\\
\omega^{-n}&0&0\\
0&0&1
\end{pmatrix}. 
\ee 
The matrix satisfies $\mathcal{C}_{\ell_n}=\mathcal{C}^\dagger_{\ell_n}=\mathcal{C}^{-1}_{\ell_n}$. 
We find that $\mathcal{C}_{\ell_n}H \mathcal{C}_{\ell_n}=H$ by using $C_{-n} M_{\ell_n}C_n=M_{\ell_n}$, and thus $\mathcal{C}_{\ell_n}$ generates the $\mathbb{Z}_2$ symmetry at $\ell_n$. 

The commutation relation between $U$ and $\mathcal{C}$ is 
\be
\mathcal{C}_{\ell_n} U\mathcal{C}_{\ell_n}=\omega^n U^{-1}, 
\label{eq:Global_Inconsistency}
\ee
and the symmetry is $(\mathbb{Z}_3)_{\mathrm{shift}}\rtimes (\mathbb{Z}_2)_{\mathcal{C}}$. 
It is important to notice that this is not the projective representation but gives the global inconsistency condition for different lines $\ell_n$~\cite{Kikuchi:2017pcp} (see also Refs.~\cite{Gaiotto:2017yup, Tanizaki:2017bam}). 
The reason this is not projective is that another $\mathbb{Z}_3$ generator $U_n=\omega^n U$ satisfies 
\be
\mathcal{C}_{\ell_n} U_n\mathcal{C}_{\ell_n}=U_n^{-1},
\ee
and thus the phase $\omega^n$ in (\ref{eq:Global_Inconsistency}) can be eliminated by redefinition of generators, which is equivalent to adding the local counter term in (\ref{eq:Global_Inconsistency_2d}). 
This, however, provides the global inconsistency condition, since the $\mathcal{C}$-symmetric state must be trivial under $U_n$, i.e. must have the $(\mathbb{Z}_3)_{\mathrm{shift}}$ charge $U=\omega^{-n}$. 
Comparing $\ell_0$ and $\ell_1$, each $\mathcal{C}$-symmetric state must have $\mathbb{Z}_3$ charge $U=1$ and $\omega^{-1}$, respectively, and these states cannot be continuously connected by $\mathbb{Z}_3$-symmetric perturbations.

%---------------------------------------------------------------------------------------
\section{Bion configuration in the twisted flag sigma model}
\label{sec:Bion}
In the previous section, we study the phase diagram of $SU(3)/U(1)^{2}$ flag sigma models with the ${\mathbb Z}_{3}$-twisted boundary condition within the dilute instanton gas approximation (DIGA), where we ignore the interaction between the fractional instantons.

The recent intensive study on the ${\mathbb C}P^{N-1}$ models with the ${\mathbb Z}_{N}$-twisted boundary condition reveals the nontrivial relation between large-order growth of perturbative series and nonperturbative contribution from fractional instanton--anti-instanton pair, called a ``bion''~\cite{Dunne:2012ae,Dunne:2012zk,Misumi:2014jua,Misumi:2014bsa,Misumi:2015dua,Fujimori:2016ljw,Fujimori:2017oab,Fujimori:2017osz,Fujimori:2018kqp}. Such a relation is one of the explicit examples of the resurgent structure, which stands for the nontrivial relation between the perturbative and nonperturbative sector: Both the perturbative Borel resummation and the nonperturbative bion contribution are accompanied with imaginary ambiguities, but they are cancelled out. All the imaginary ambiguities are cancelled as we incorporate all the multi-bion contributions, then we end up with the real physical quantities without ambiguities. 

We speculate a similar structure in the $SU(3)/U(1)^{2}$ flag sigma models with ${\mathbb Z}_{3}$-twisted boundary condition too. In the following, toward future intensive investigation on the resurgent structure of the theory, we show existence of the imaginary ambiguity in the contributions from the bion.

\begin{figure}[t]
 \centering
 \includegraphics[scale=0.34]{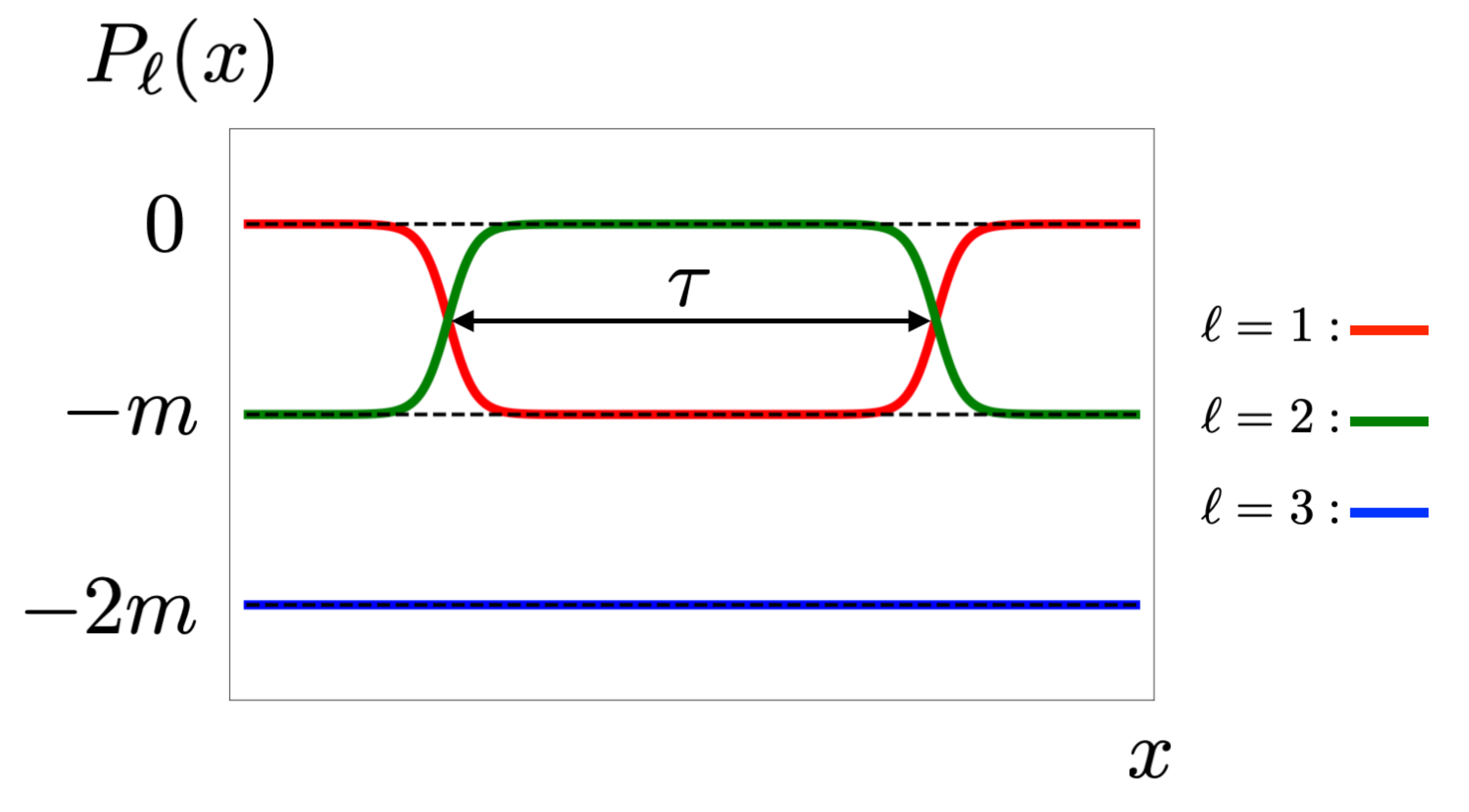}\\
 \caption{
 Bion configuration in $SU(3)/U(1)^{2}$ flag sigma models with ${\mathbb Z}_{3}$-twisted boundary condition.
 }
 \label{fig:Bion}
\end{figure}

The bion amplitude in the $SU(N)/U(1)^{N-1}$ flag sigma models on ${\mathbb R}\times S^{1}$ with the ${\mathbb Z}_{N}$-twisted boundary condition is calculated via quasi-moduli integral of the interaction potential between fractional instanton and fractional anti-instanton.
One of the bion configuration is given by
\begin{equation}
 \begin{split}
 \bphi_1^{\calI \bar{\calI}} &= { \rme^{(z-\overline{z})/2} \over \sqrt{1+|\rme^{z-z_1} + \rme^{-\overline{z}+\overline{z}_{2}}|^2}}
 \begin{pmatrix}
 \rme^{\overline{z}-\overline{z}_1} + \rme^{-z+z_{2}} \\
 -1 \\
   0 
 \end{pmatrix},\,
  \\
  \bphi_2^{\calI \bar{\calI}} &=   {1\over \sqrt{1+|\rme^{z-z_1} + \rme^{-\overline{z}+\overline{z}_{2}}|^2} }
 \begin{pmatrix}
   1 \\
 \rme^{z-z_1} + \rme^{-\overline{z}+\overline{z}_{2}} \\
   0
 \end{pmatrix},\,  
  \\
 \bphi_3^{\calI \bar{\calI}} &= 
 \begin{pmatrix}
     0 \\
     0 \\
      \rme^{-{2\pi \im\over 3L} t}
 \end{pmatrix}, 
 \end{split}
  \label{eq:bionconf}
\end{equation}
where the parameters 
\begin{align}
z_{1}\equiv {2\pi \over{3L}}x_{1} +\im \phi_{1},\,\quad\quad
z_{2}\equiv {2\pi \over{3L}}x_{2} +\im \phi_{2},
\end{align}
include two moduli composed of a center position and an overall phase and two quasi-moduli composed of a relative distance and a relative phase. Among them we here focus on the quasi-moduli parameters, which are not genuine moduli but come to be moduli in the well-separated limit (or in the weak-coupling limit in the complexified theory~\cite{Fujimori:2016ljw}). They are given by
\begin{equation}
\chi  \equiv {x_{1}-x_{2} \over{L}},\quad\quad
\phi\equiv \phi_{1}-\phi_{2}\,,
\label{eq:qmoduli}
\end{equation}
with $-\infty<\chi<\infty$ and $-\pi\leq\phi<\pi$.
Here we define both quasi-moduli $\chi, \phi$ as dimensionless parameters.
Eq.~\eqref{eq:bionconf} is regarded as an approximate configuration of the complex bion solutions in the complexified theory~\cite{Behtash:2015zha,Fujimori:2016ljw,Fujimori:2017oab,Fujimori:2017osz,Fujimori:2018kqp}.
We schematically depict the bion configuration in $SU(3)/U(1)^{2}$ flag sigma models with the 
${\mathbb Z}_{3}$-twisted boundary condition in Fig.~\ref{fig:Bion}.

Let us perform the quasi-moduli integral although we need the one-loop determinant around the bion background to obtain the whole single-bion contribution to the partition function ${Z_{\calI\bar{\calI}} \over{Z_{0}}}$.
The (bare) bion effective potential in the $SU(3)/U(1)^{2}$ flag sigma model on ${\mathbb R}\times S^{1}$ with the ${\mathbb Z}_{3}$-twisted boundary condition is derived by substituting the bion configuration \eqref{eq:bionconf} into the action. It is notable that, as a result of the constraints on the field variable in the flag sigma model, the number of quasi-moduli parameters in the bion configuration of the $SU(3)/U(1)^{2}$ flag sigma model becomes two as shown in eq.~\eqref{eq:qmoduli}, which coincides with that of the ${\mathbb C}P^{2}$ sigma model. Eventually, the bion contribution to the partition function in the present model is obtained from the bion effective action equivalent to twice of that in the ${\mathbb C}P^{2}$ sigma model \cite{Fujimori:2016ljw} as
\begin{align}
{Z_{\calI\bar{\calI}} \over{Z_{0}}} &\,=\, {\mathcal K} L
\int_{-\pi}^{\pi} \diff \phi \int_{-\infty}^{\infty} \diff \chi \, \rme^{-V(\chi,\phi)},
\\
V(\chi,\phi) &\,=\,  2V_{{\mathbb C}P^{2}}(\chi,\phi)\,=\,
{2m\over{g}}-{4m\over{g}}\cos\phi 
\,\rme^{-m \chi}\, + 4m \epsilon \chi,
\label{potNB}
\end{align}
with $m\equiv 2\pi/3$. Here all the parameters and coordinates are made dimensionless as $m\equiv {2\pi\over{3L}} L$ and $\chi=(x_{1}-x_{2})/L$ by the use of the compactification circumference $L$, and this is why the $L$ appears as an overall factor in ${Z_{\calI\bar{\calI}} \over{Z_{0}}}$. 
The factor ${\mathcal K}$ includes contributions from one-loop determinant and the genuine moduli integral over a center position and an overall phase.
We note that the definition of coupling here is obtained by replacing as $2/g^2 \to 1/g$ in the definition of the references~\cite{Fujimori:2016ljw,Fujimori:2018kqp}.
The last term $4m\epsilon \chi$ corresponds to the deformation term originating in quantum-mechanical fermionic degrees of freedom. We are here interested only in the bosonic flag sigma model, thus we take a $\epsilon\to 0$ limit in the end of calculation.
Precisely speaking, this bion effective action will be renormalized by taking account of the Kaluza-Klein modes of quantum fluctuations around the bion configuration, where the coupling will be renormalized~\cite{Fujimori:2018kqp}. However, we here concentrate on this bare bion effective action to show the existence of the imaginary ambiguity.

It is shown in Ref.~\cite{Fujimori:2016ljw} that this integral is performed by complexifying $\chi, \phi$ and applying the Lefschetz thimble decomposition of the integration contour~\cite{Witten:2010cx, Witten:2010zr, Cristoforetti:2012su, Cristoforetti:2013wha, Fujii:2013sra, Tanizaki:2014xba, Cherman:2014sba, Tanizaki:2014tua, Kanazawa:2014qma, Tanizaki:2015pua, DiRenzo:2015foa, Fukushima:2015qza, Tanizaki:2015rda, Fujii:2015bua, Fujii:2015vha, Alexandru:2015xva, Alexandru:2015sua, Tanizaki:2016xcu, Tanizaki:2017yow}, which corresponds to the thimble decomposition of the complexified path integral associated with complex and real bion saddle points.
We here calculate it in a distinct but equivalent manner following Ref.~\cite{Misumi:2015dua}.
We first consider the integrals
\begin{align}
\quad\quad\quad\quad
I\left({1\over{g}} \right)&\equiv
\int_{-\infty}^\infty \diff \chi \exp
\left(
{4m \over{g}} \rme^{-m \chi}
-4m \epsilon \chi
\right)
%\nonumber \\ 
%&=
%{1\over{m}}\int_{0}^{\infty} \diff s\,\rme^{-s}s^{4\epsilon-1} \left({-g \over{4m}}\right)^{4\epsilon} %\nonumber\\
\,=\,
{1\over{m}}\Gamma(4\epsilon) \left({-g \over{4m}}\right)^{4\epsilon} 
%\quad\quad\quad\quad\quad \left(s\equiv {4m\over{-g}}\rme^{-m \chi}\right)
\nonumber\\
%&=-{1\over{m}}\left(\gamma +\log \frac{4m}{\rme^{\mp \im\pi} g}\right)+{\mathcal O}(\epsilon) +  {\mathcal O}\left({1\over\epsilon}\right)\nonumber \\ 
&=
-{1\over{m}}\left(\gamma +\log \frac{4m}{g}  \mp \im\pi \right)
+{\mathcal O}(\epsilon) +  {\mathcal O}\left({1\over\epsilon}\right),\quad\quad\quad\quad
 \left(-g = \rme^{\pm \im \pi} g\right)
\label{eq:bion_no_phase}
\end{align}
and
\begin{align}
I\left(-{1\over{g}} \right)&\equiv
\int_{-\infty}^\infty \diff \chi \exp
\left(-
{4m \over{g}}\rme^{-m \chi}
-4m \epsilon \chi
\right)
%\nonumber \\ 
%&=
%{1\over{m}}\int_{0}^{\infty} \diff s\,\rme^{-s} s^{4\epsilon-1} \left({g \over{4m}}\right)^{4\epsilon} 
%\nonumber\\
\,=\,
{1\over{m}}\Gamma(4\epsilon) \left({g \over{4m}}\right)^{4\epsilon} 
%\quad\quad\quad\quad\quad \left(s\equiv {4m\over{g}}\rme^{-m \chi}\right)
\nonumber\\
&=
-{1\over{m}}\left(\gamma +\log \frac{4m}{g}\right)
+{\mathcal O}(\epsilon) +  {\mathcal O}\left({1\over\epsilon}\right)\,.
\label{eq:bion_no_phase}
\end{align}
In the former integral we come across the imaginary ambiguity since we first need to regard $-g$ as positive-valued and take analytic continuation as $-g = \rme^{\pm \im \pi} g$ in the end.
This procedure of the integral is called the Bogomol'nyi--Zinn-Justin prescription~\cite{ZinnJustin:1981dx}.
On the other hand, the imaginary ambiguity is absent when the argument of $I(X)$ is negative as in the latter integral since we need no analytic continuation.
With taking account of the relative phase moduli $\phi$, the bion contribution to the partition function is expressed as
\begin{equation}
{Z_{\calI\bar{\calI}} \over{Z_{0}}}= {\mathcal K} \,L\,\rme^{-{4\pi\over{3g}}}\int_{-\pi}^{\pi} \diff \phi \,I\left({\cos\phi \over g}
\right), 
\end{equation}
where $4\pi/(3g)=2S_{\calI}$ is twice of the fractional instanton action of the present theory. 
Since we apply the above Bogomol'nyi--Zinn-Justin prescription to the integral, 
we need to decompose the integration region into two, where the arguments of $I(X)$ are positive and negative respectively.
The main part of the integral is then performed as
\begin{align}
m\int_{-\pi}^{\pi} \diff \phi\,
I\left({\cos\phi \over g}
\right)
&=
2m
\int_{0}^{\pi/2} 
\diff \phi\,
I\left({\cos\phi \over g}\right)
%\nonumber\\
%&&\quad\quad
+
2m
\int_{\pi/2}^{\pi}\diff \phi \,
I\left({\cos\phi \over g}\right)
%\nonumber\\&=
%-\pi\left(\gamma +\log \frac{4m}{g}- \log2\right) \mp \im\pi^{2}
%\nonumber\\
%&\quad\quad
%-\pi \left(\gamma +\log \frac{4m}{g}- \log2\right) 
\nonumber \\
&=
-2\pi \left(\gamma +\log \frac{2m }{g}\right)
\,\mp\, \im \pi^{2}  \,\,\,
+{\mathcal O}(\epsilon) + {\mathcal O}\left({1\over \epsilon}\right),
\end{align}
where we used $\int_{0}^{\pi/2}\diff \phi \log(\cos \phi)=-{\pi\over{2}}\log 2$.
In the whole partition function, the ${\mathcal O}(1/\epsilon)$ term is expected to be cancelled by that in the other quasi-moduli integral, thus we drop this here.
We thus obtain the bion contribution from eq.~(\ref{eq:bionconf}) in the $\epsilon\to0$ limit as 
\begin{equation}
{Z_{{\calI}\bar{\calI}} \over{Z_{0}}}
\,=\, 
%Ce^{-\frac{2S_{I}}{N}}\left[-2\pi\left(\gamma +\log (2\kappa L v^2) \right)
%\,\mp\, i\pi^{2} \right]= 
e^{-\frac{4\pi}{3g}} {\mathcal K} {L\over{m}}
\left[-2\pi\left(\gamma +\log \frac{2m }{g}\right)
\,\mp\, i \pi^{2}  \right]\,,
\label{E0NB}
\end{equation}
with $m=2\pi/3$.
The bion contribution to ground-state energy from (\ref{eq:bionconf}) is obtained as $E_{{\calI}\bar{\calI}} =-\lim_{\beta\to\infty}{1\over{\beta}}{Z_{{\calI}\bar{\calI}} \over{Z_{0}}}$, where $\beta$ is the size of the $x$-direction introduced to regularize the partition function. 
Although the factor ${\mathcal K}$ is not calculated here, we can speculate the behavior of the perturbation series by assuming the resurgent structure, where the imaginary ambiguities arising from the perturbative Borel resummation and the bion contribution are cancelled out.
From eq.~\eqref{E0NB}, the Borel transform of perturbative series of the ground-state energy is expected to have a Borel singularity at $t=4\pi/3$ when the perturbative parameter is $g$.

We note that the Borel singularity position does not coincide with that of the infrared-renormalon singularity expected from the coefficient of the beta function in eqs.~\eqref{eq:RG}-\eqref{eq:RG_running}.
As we have mentioned, however, the bion effective action for the finite compactification circumference is renormalized by summing over the Kaluza-Klein modes of quantum fluctuations around bion configurations as shown in \cite{Fujimori:2018kqp}. It is an interesting question whether or not we can show that the bion contribution includes the imaginary ambiguity corresponding to the infrared-renormalon as with the case in the ${\mathbb C}P^{N-1}$ model \cite{Fujimori:2018kqp} by taking account of the quantum fluctuations. These topics are left for future works.

%--------------------------------------------------------------------------------------
\section{Summary and Discussion}
\label{sec:Summary}

In this work, we studied the vacuum structure and the phase diagram of the $SU(3)/U(1)^{2}$ flag sigma model on $\mathbb{R} \times S^1$ with the ${\mathbb Z}_{3}$-twisted boundary condition.
The phase diagram in the $(\theta_{1},\theta_{2})$-plane is obtained from the dilute instanton gas approximation (DIGA) at small compactification radius, and it is consistent with the one conjectured for the uncompactified theory. 
This indicates the adiabatic continuity of this $2$-dimensional asymptotically-free theory.

What we have done in the present work is summarized as follows:
\begin{itemize}
\item
We classified the classical vacua and derived the fractional instanton solutions connecting those vacua. 
%We found the special property of the fractional instantons that the total topological charge is zero. 
\item
We obtained the eigenenergies within the dilute instanton gas approximation. 
The phase diagram with respect to the two theta angles $(\theta_{1},\theta_{2})$ shows good agreement with the conjectured one for the uncompactified theory.
\item
We computed contributions from the instanton--anti-instanton configuration, called a bion, and showed the existence of the imaginary ambiguity. This ambiguity is expected to be cancelled by that from the perturbative Borel resummation. 
\end{itemize}

Although our study on the phase diagram by the use of the DIGA is restricted to the $SU(3)/U(1)^{2}$ flag sigma model in this work, it can be extended to general $SU(N)/U(1)^{N-1}$ flag sigma models as long as one can find fractional instanton solutions for ${\mathbb Z}_{N}$-twisted boundary conditions. 
%It is of great interest to study whether the topological charge of the fractional instantons are also zero for the $SU(N)/U(1)^{N-1}$ ($N\geq 4$) flag sigma models on $\mathbb{R}\times S^1$ with the ${\mathbb Z}_{N}$ twisted boundary condition. 
We speculate that we will find a similar sign of the adiabatic continuity in these general flag sigma models too. 

One important physics that could not be differentiated within the DIGA is whether the triple degeneracy at $(\theta_1,\theta_2)=(2\pi/3,-2\pi/3)$ of our model comes out of the two-dimensional conformal behavior or not. Since the compactification sets the energy scale $L^{-1}$, both the two-dimensional conformal behavior and spontaneous $\mathbb{Z}_3$ breaking can explain the triple degeneracy in the ground states of $SU(3)/U(1)^2$ sigma model on $\mathbb{R}\times S^1$. 
So far, this problem was investigated in two-dimensions using the numerical analytic continuation of the Monte Carlo results with the imaginary theta angles~\cite{Lajko:2017wif}. 
It is very desirable to get the reliable result from numerical simulations with real theta angles, and we here point out that this recently becomes possible by lattice dual formulation~\cite{Gattringer:2018dlw}. 
We also have a chance to get new insight by considering the supersymmetric version of the flag sigma model, which was introduced as marginal deformation of supersymmetric WZW model~\cite{Israel:2004cd,Israel:2004vv}. 

Investigation on the resurgent structure of the flag sigma model is also a theme to be studied in details. The quantum fluctuations around the bion configurations (or complex bion solutions) for the $2$-dimensional ${\mathbb C}P^{N-1}$ sigma model on $\mathbb{R} \times S^1$ with the ${\mathbb Z}_{3}$-twisted boundary condition has been recently calculated in \cite{Fujimori:2018kqp} by summing over the Kaluza-Klein modes. It was there shown that the bion effective action is renormalized and the renormalized coupling (or the dynamical scale) emerges correctly. The bion contribution obtained from this renormalized effective bion action yields the imaginary ambiguity consistent with the infrared renormalon ambiguity arising from the perturbative Borel resummation. This procedure for verifying the resurgent structure can be extended to the flag sigma models.

Let us also propose a possible interesting connection 
between the fractional instantons of the $SU(N)/U(1)^{N-1}$ flag sigma models and the nontrivial saddles of the $SU(N)$ principle chiral model (PCM).
In Ref.~\cite{Tanizaki:2018xto}, level-$p$ $SU(N)$ Wess-Zumino-Witten (WZW) model is continuously deformed to $SU(N)/U(1)^{N-1}$ flag sigma model with $\theta_\ell=2\pi p\ell/N$ by adding the double-trace term. 
Since the classical action of the level-$0$ WZW model coincides with that of the $SU(N)$ PCM, this suggests possible connection between $SU(N)/U(1)^{N-1}$ flag sigma models and $SU(N)$ PCM. What is specific to PCM is that we cannot define topological charge since $\pi_{2}[SU(N)]=0$. Nevertheless, it has been shown that there are solutions similar to fractional instantons called Uhlenbeck's ``fractons'' and ``unitons'' in PCM on $\mathbb{R}\times S^1$, which of course has no topological charge \cite{Piette:1987qp,Piette:1987qr,Cherman:2013yfa}.
We speculate that the fractional instantons for the flag sigma model correspond to these nonperturbative configurations in PCM. If this observation is correct, the DIGA we have done in this work may be interpreted as ``dilute fracton (uniton) gas approximation'' in PCM. Further investigation is required for revealing this possible correspondence.

In the end of this paper, we comment on the extension of the study to a broad range of sigma models.
The flag sigma models in a broad sense includes ${\mathbb C}P^{N-1}$, Grassmannian, flag sigma models in a narrow sense and more. The study based on combination of 't Hooft anomaly matching and semiclassical analyses including the investigation on the resurgent structure can be extended to such sigma models, where we expect to gain fruitful outcomes.

%-------------------------------------------------------------------------------------
\acknowledgments
M.~H. is supported by the Special Postdoctoral Researchers Program and iTHEMS Program (iTHEMS STAMP working group) at RIKEN. 
Y.~T. is supported by Special Postdoctoral Researchers Program of RIKEN. 
T.~M. is supported by the Japan Society for the Promotion of Science (JSPS) 
Grant-in-Aid for Scientific Research (KAKENHI) Grant Numbers 
16K17677 and 18H01217.
T.~M. is also supported by the Ministry of Education, Culture, 
Sports, Science, and Technology(MEXT)-Supported Program for the 
Strategic Research Foundation at Private Universities ``Topological 
Science'' (Grant No. S1511006). 
%-------------------------------------------------------------------------------------

\appendix

\section{Fractional instantons of $\mathbb{C}P^{N-1}$ sigma model}
\label{sec:CPNFracInstanton}

The $\mathbb{C}P^{N-1}$ model is defined by 
\be
S={1\over 2g}\int |D\phi|^2+{\im \theta\over 2\pi}\int \diff a. 
\ee
Here, $\phi$ is $N$-component complex fields with $|\phi|^2=1$, $D\phi=(\diff+\im a)\phi$ is the covariant derivative, and $a$ is $U(1)$ gauge field. By solving the equation of motion, we get 
\be
a=\im \overline{\phi}\cdot\diff \phi. 
\ee

To obtain the (anti-)self-dual equation, let us rewrite the kinetic term as 
\bea
\int \overline{D\phi}\wedge \star D \phi&=&{1\over 2}\int \overline{(D\phi\pm \im \star D\phi)}\wedge \star (D\phi\pm \im  \star D\phi)\pm \im\int \overline{D\phi}\wedge D \phi\nonumber\\
&=&{1\over 2}\int |D\phi \pm \im \star D\phi|^2\pm \int \diff a. 
\eea
Therefore, for a fixed topological charge, the minimal action is given by the (anti-)BPS solution~\cite{Bogomolny:1975de, Prasad:1975kr},
\be
D\phi\pm \im \star D\phi=0. 
\ee
Introducing the stereographic coordinate, $\phi$ is represented by the $(N-1)$-component complex field $n$ as
\be
\phi={1\over \sqrt{1+n^\dagger n}}\begin{pmatrix}
1\\n
\end{pmatrix}. 
\ee
Then, the (anti-)BPS equation becomes the (anti-)holomorphic condition, 
\be
(\p_x\pm \im \p_t)n=0. 
\ee

The fractional instanton appears by introducing the twisted boundary condition on $\mathbb{R}\times S^1$, 
\be
\phi(x,t+L)=C\phi(x,t),
\ee
where $C$ is the clock matrix, $\mathrm{diag}(1,\omega,\ldots, \omega^{N-1})$, with $\omega=\rme^{2\pi\im/N}$. 
Using the dimensionless complex coordinate $z={2\pi\over NL}(x+\im t)$, the fractional instanton is given by 
\be
\phi={1\over \sqrt{1+|\rme^{z-z_0}|^2}}
\begin{pmatrix}
1\\
\rme^{z-z_0}\\
0\\
\vdots\\
0
\end{pmatrix}, 
\ee
where $z_0$ is the moduli parameter. The $x$-dependence of the Polyakov-loop phase is
\be
\int_{S^1}a=\int_0^L\diff t{\im\over (1+n^\dagger n)}n^\dagger \p_t n=-{2\pi\over N}{1\over 1+\rme^{-{4\pi\over NL}(x-x_0)}}. 
\ee
Therefore, the topological charge is given by $-2\pi/N$, which is why the configuration is called fractional instanton.

\bibliographystyle{utphys}
\bibliography{./QFT,./refs,./lefschetz}
%\bibliography{./Lef_thimble}
\end{document}